\documentclass[
a4paper,
aps,
prd,
twocolumn,
tightenlines,
preprintnumbers,
nofootinbib,
showkeys
]{revtex4-1}

\usepackage{XCharter}
\usepackage[T1]{fontenc}

\usepackage{fullpage}
\usepackage{amsfonts}
\usepackage{amsmath}
\usepackage{slashed}
\usepackage{amssymb}
\usepackage{graphicx}
\usepackage{epic}
\usepackage{eepic}
\usepackage{epsfig}
\usepackage{latexsym}
\usepackage{color}
\usepackage{float}
\usepackage{multirow}
\usepackage{hyperref}
\usepackage{enumitem}
\hypersetup{colorlinks=true,citecolor=red,linkcolor=magenta,urlcolor=blue}
\usepackage[caption=false]{subfig}
\usepackage{natbib}
\usepackage{relsize}
\usepackage[left=2cm,right=2cm,top=1.9cm,bottom=1.95cm]{geometry}

\usepackage{shorthand}
\RequirePackage{times}

\newcommand{\ubr}[1]{\raisebox{1.5ex}{\hspace{#1ex}$\frown$\relax}}

\begin{document}
\relscale{0.95}

\title{$R_{D^{(*)}}$ motivated $\mc S_1$ leptoquark scenarios: Impact of interference on the exclusion limits from LHC data}

\author{Tanumoy Mandal}
\email{tanumoy.mandal@physics.uu.se}
\affiliation{Department of Physics and Astrophysics, University of Delhi, Delhi 110007, India}
\affiliation{\vspace{-2.5ex}Department of Physics and Astronomy, Uppsala University, Box 516, SE-751 20 Uppsala, Sweden}

\author{Subhadip Mitra}
\email{subhadip.mitra@iiit.ac.in}
\affiliation{Center for Computational Natural Sciences and Bioinformatics,
International Institute of Information Technology, Hyderabad 500 032, India}

\author{Swapnil Raz}
\email{swapnil.raz@research.iiit.ac.in}
\affiliation{Center for Computational Natural Sciences and Bioinformatics,
International Institute of Information Technology, Hyderabad 500 032, India}
\date{\today}

\begin{abstract}
\noindent 
Motivated by the persistent anomalies in the semileptonic $B$-meson decays, we investigate the competency of 
LHC data to constrain the $R_{D^{(*)}}$-favoured parameter space in a charge 
$-1/3$ scalar leptoquark ($\mathcal S_1$) model. We consider some scenarios with one large free coupling
to accommodate the 
$R_{D^{(*)}}$ anomalies. As a result, some of them dominantly yield nonresonant $\tau\tau$ and $\tau\nu$ events 
at the LHC through the $t$-channel $\mathcal S_1$ exchange. So far, no experiment has searched for leptoquarks using 
these signatures and the relevant resonant leptoquark searches are yet to put any strong exclusion limit on the 
parameter space. We recast the latest $\tau\tau$ and $\tau\nu$ resonance search data to obtain new exclusion 
limits. The nonresonant processes strongly interfere (destructively in our case) with the Standard 
Model background and play the determining role in setting the exclusion limits. To obtain precise limits, we include 
non-negligible effects coming from the subdominant (resonant) pair and inclusive single leptoquark productions 
systematically in our analysis. To deal with large destructive interference, we make use of the transverse mass 
distributions from the experiments in our statistical analysis. In addition, we also recast the relevant direct search 
results to obtain the most stringent collider bounds on these scenarios to date. These are independent bounds and are competitive to other known bounds. Finally, we indicate how one can adopt these bounds 
to a wide class of models with $\mathcal S_1$ that are proposed to accommodate the $R_{D^{(*)}}$ anomalies.
\end{abstract}


\maketitle

\section{Introduction}\label{sec:intro}
\noindent
The Standard Model (SM) is known to describe the interactions among the elementary particles extremely well -- 
it has been spectacularly successful in its predictions. However, there are several theoretical as well as experimental 
reasons to believe that the SM is not the ultimate theory, rather, it is an effective theory of the sub-TeV energy scales. 
Motivated by the new physics models proposed to address some unexplained issues in the SM, one normally expects at 
the TeV energy scale, some new interactions and/or particles would be visible. Because of this, after the discovery of 
the Higgs boson, signatures of physics beyond the Standard Model (BSM) are being searched for  extensively at the 
Large Hadron Collider (LHC). 

The direct detection searches for new physics at the CMS and the ATLAS detectors of the LHC have not found any 
evidence so far. But some really intriguing hints towards new physics have been observed repeatedly by different 
experiments in some $B$-meson decays that violate lepton flavour universality. The most drastic departure from the 
SM expectation was first noticed by the BaBar collaboration in 2012 \cite{Lees:2012xj, Lees:2013uzd}. They reported 
an excess of about 3.4$\sigma$ in the ratio of $B$-meson semileptonic decay branching fractions,
\ba
R_{D^{(*)}}=\frac{Br(B \to D^{(*)}\tau\nu)}{Br(B \to D^{(*)}\ell\nu)} {\rm~where ~} \ell=\{e,\mu\},\quad
\ea
than the SM expectation. Their results were consistent with the measurements by the Belle collaboration 
\cite{Matyja:2007kt,Adachi:2009qg,Bozek:2010xy, Huschle:2015rga,Sato:2016svk,Hirose:2016wfn}. 
Later LHCb also confirmed this anomaly for $\displaystyle R_{D^*}$\cite{Aaij:2015yra,Aaij:2017deq} 
(see Table~\ref{tab:rdstarcomparison} for a comparison of the different results). 
\begin{table*}[t]\vspace{-1.5ex}
\caption{Values of  $R_{D^{(*)}}$: The SM predictions, the results from different experiments and the 
averages of the experimental results obtained by the Heavy Flavor Averaging Group (HFLAV). For ease, 
the statistical and systematic errors in the experimental numbers have been added in quadrature.}\label{tab:rdstarcomparison}
\begin{tabular}{| c | c | c | c  | c |}\hline
SM&	BaBar&	Belle&	LHCb&	HFLAV Avgs.\cite{Amhis:2016xyh}\\\hline\hline
\multicolumn{5}{|l|}{$R_{D^*}$ results} \\\hline
&	&	$0.293 \pm 0.041$~\cite{Huschle:2015rga}&	$0.336$ $\pm$ $0.040$~\cite{Aaij:2015yra}&\\
$0.252 \pm 0.004$~\cite{Tanaka:2012nw} &$0.332 \pm 0.030$~\cite{Lees:2012xj}&	$0.302 \pm 0.032$~\cite{Sato:2016svk}&$0.291 \pm 0.035$~\cite{Aaij:2017deq}&$0.306 \pm 0.015$ \\
&&	$0.270^{\,+\,0.045}_{\,-\,0.043}$~\cite{Hirose:2016wfn}&&\\\hline
\multicolumn{5}{|l|}{$R_{D}$ results} \\\hline
$0.299 \pm 0.011$~\cite{Lattice:2015rga}& $0.440 \pm 0.072$~\cite{Lees:2012xj}& $0.375 \pm 0.069$~\cite{Huschle:2015rga}&\hfill-\hfill~&$0.407 \pm 0.046$\\\hline
\end{tabular}
\end{table*}
Together, these measurements 
amount to about a 4$\sigma$ deviation from the SM expectation \cite{Freytsis:2015qca}.
Another anomaly was recently reported by the LHCb collaboration in the $B$-meson leptonic decays~\cite{Aaij:2014ora,Aaij:2017vbb}. They 
measured the following ratio,  
\ba
R_{K^{(*)}}=\frac{Br(B \to K^{(*)}\mu^+\mu^-)}{Br(B \to K^{(*)}e^+e^-)}.
\ea 
They obtained values that are about $2.5\sigma$ smaller than the corresponding SM estimations \cite{Hiller:2003js,Bordone:2016gaq}.

\begin{figure*}\vspace{-1.5ex}
\subfloat[]{\includegraphics[width=0.4\textwidth]{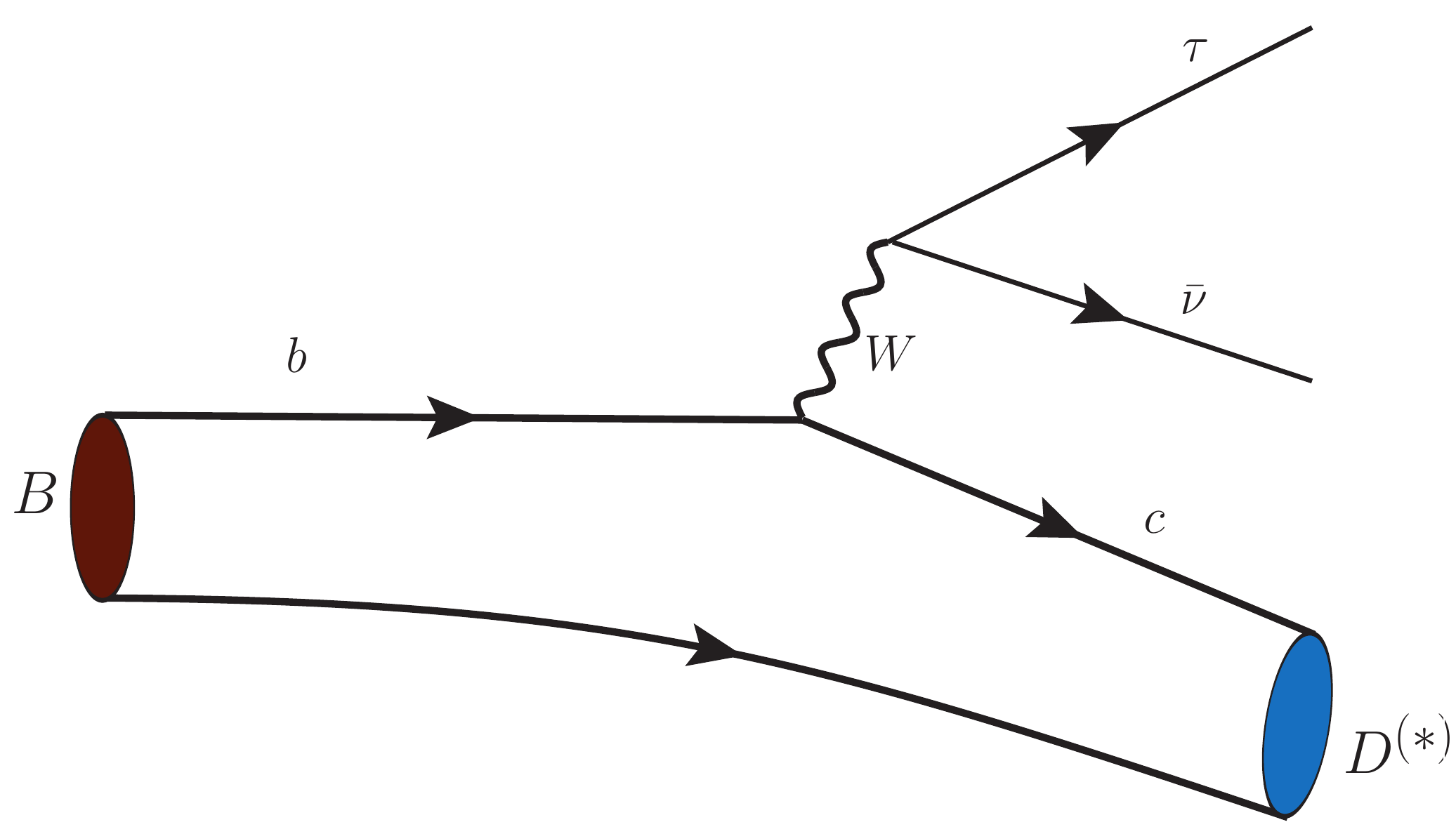}\label{fig:BtoDa}}\hspace{2cm}
\subfloat[]{\includegraphics[width=0.4\textwidth]{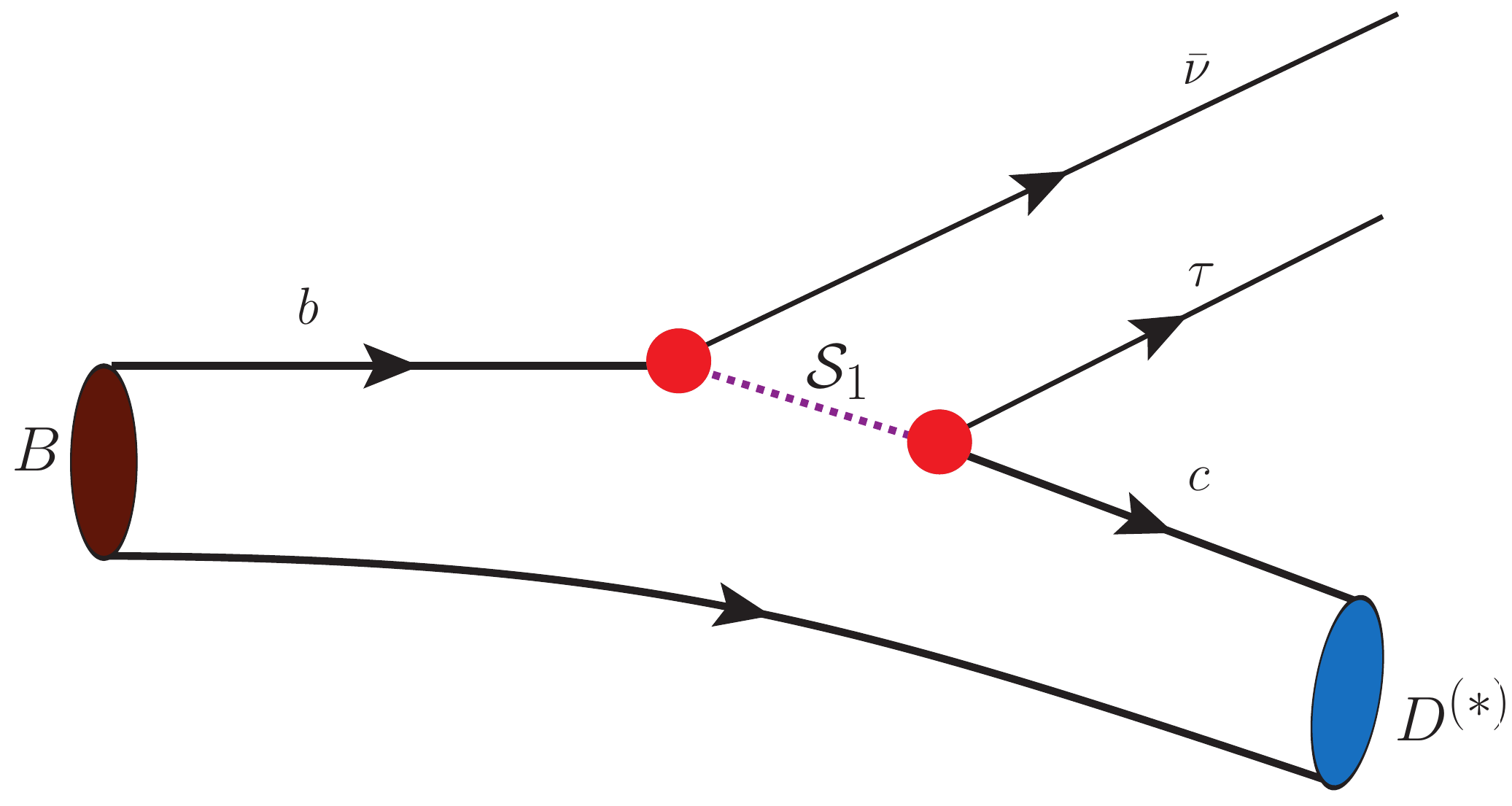}\label{fig:BtoDb}}
\caption{Leading order processes responsible for $B\to {D^{(*)}}\tau\nu$ decay. In the SM (a) and in the $\mc S_1$ leptoquark model (b).}\label{fig:BtoD}
\end{figure*}

In the literature, several proposals have been put forward to address these anomalies. In this paper, we look at 
the $R_{D^{(*)}}$ anomalies. At the leading order in the SM, the semileptonic ${B\to D^{(*)}}$ decays proceed 
through a $b\to cW$ transition with the $W$ boson decaying further to a charged lepton and a neutrino 
[see e.g., Fig.\ref{fig:BtoDa}]. Since the experiments are indicating towards an enhanced $\tau$-mode, any new 
physics model that can contribute positively to the $b\to c\tau\nu$ decay could accommodate a possible explanation 
as long as it does not predict a similar enhancement to the $\ell\nu$-modes. Among various proposals, leptoquark 
(LQ, $\ell_q$) explanations have received a lot of attention in the literature (see e.g. 
\cite{Dorsner:2013tla,Sakaki:2013bfa,Bauer:2015knc,Becirevic:2016yqi,Sahoo:2016pet,Hiller:2016kry,Crivellin:2017zlb,Cai:2017wry,Assad:2017iib, Becirevic:2018afm,Angelescu:2018tyl,Bansal:2018nwp}). 
LQs are hypothetical colour-triplet bosons (scalar or vector) that also carry nonzero lepton and baryon quantum 
numbers. Hence, a LQ can couple to a lepton and a quark and has fractional electromagnetic charge. Since the 
$b\to c\tau\nu$ process involves two quarks and two leptons, LQs that couple to these fermions could be a 
good candidate to explain the $R_{D^{(*)}}$ anomalies [see e.g., Fig.~\ref{fig:BtoDb}].  

LQs are an important ingredient in many BSM theories. For example, they appear in different BSM scenarios like 
Pati-Salam  models \cite{Pati:1974yy}, the models with quark lepton compositeness \cite{Schrempp:1984nj}, 
$\mathrm{SU}(5)$ grand unified theories \cite{Georgi:1974sy}, $R$-parity violating supersymmetric models 
\cite{Barbier:2004ez} or coloured Zee-Babu model \cite{Kohda:2012sr} etc. Their phenomenology has also 
been studied in great detail (see, for example, Refs.~\cite{Arnold:2013cva, Bandyopadhyay:2018syt, Vignaroli:2018lpq} 
for some phenomenological studies).

There are many models with a single LQ (with various quantum numbers) that have been discussed in the context 
of heavy flavour anomalies (see e.g., Refs.~\cite{Hiller:2016kry,Cai:2017wry,Angelescu:2018tyl} for an overview). 
In this paper, we study the current LHC bounds on a simple model with only one LQ that could accommodate the $R_{D^{(*)}}$ 
anomalies. Our aim is to  investigate whether the current LHC data alone can constrain the $R_{D^{(*)}}$-favoured parameter 
space in this model. Our approach here is a phenomenologically motivated bottom-up one.  For explaining the $R_{D^{(*)}}$ 
anomalies, a LQ that couples to the third generation lepton(s) and, second and third generation quarks ($c$ and $b$) is 
required [see Fig.~\ref{fig:BtoDb}]. Here, for simplicity, we consider 
a model that has one scalar LQ that is weak singlet and has electromagnetic charge $\displaystyle -1/3$. This type of 
LQ is commonly denoted as $\mc S_1$~\cite{Buchmuller:1986zs} (also as $\mc S_0$~\cite{Dorsner:2016wpm}). We postpone 
similar analysis for other possible LQs to a future publication.

Earlier, it has been shown that to resolve the $R_{D^{(*)}}$ anomalies with $\mc S_1$, one generally introduces some 
large new coupling(s) that would affect other flavour observables or precision electroweak tests bounds (see 
e.g., Refs.~\cite{Hiller:2016kry,Cai:2017wry,Angelescu:2018tyl,Bansal:2018nwp}). Here, however, we do not discuss these 
bounds. Instead, our aim is to obtain complimentary limits from LHC data that are independent of the other bounds. 
Generally, it may be possible to avoid some model specific bounds by introducing new degree(s) of freedom in the theory (like 
Ref.~\cite{Crivellin:2017zlb} shows how one can make a model of $\mc S_1$ consistent with the bound on $b\to s\nu\nu$ 
by introducing another triplet scalar LQ). As we shall see,  one has to make some minimal assumptions 
about the model to obtain the LHC bounds. But, once the minimal assumptions are satisfied, it is not possible to completely  bypass the LHC bounds simply by extending the model.   
To obtain the LHC bounds, we consider some minimal scenarios where the 
model depends only on one new parameter (coupling) that becomes relevant for the $R_{D^{(*)}}$ observables (apart from 
$M_{\mc S_1}$, the mass of $\mc S_1$). In this simple setup, it is possible to obtain constraints on this parameter from the 
experiments in a straightforward manner. One can then use them as templates for obtaining bounds on complex setups with 
more degrees of freedom.

In this paper, we study two  
minimal scenarios. Of these, the LHC phenomenology of one has not been explored in detail earlier; the direct detection bounds on 
them are weak. Here, we recast the LHC dilepton and monolepton+$\slashed{E}_{\rm T}$ search results. We find that these 
searches have already put severe constraints on the new coupling in this scenario. In the other minimal scenario, a different 
new coupling is present but the LHC is mostly insensitive to it. In this case, the only available bounds are on the mass of the 
LQ from the direct detection searches. For completeness, we also study the bounds in an intermediate next-to-minimal scenario 
where both of these new couplings are nonzero.  

Before we proceed further, we review the direct detection bounds on LQs that couple with third generation fermions available 
from the LHC. Assuming $Br(\ell_{q}\to t \tau)=1$, a recent scalar LQ pair production search at the CMS detector has excluded masses 
below $900$ GeV \cite{Sirunyan:2018nkj}. Also, reinterpreting their search results for squarks and gluinos, CMS has put bounds 
on both the scalar and vector LQs that decay to a third generation quark and a neutrino~\cite{Sirunyan:2018kzh}. For a scalar 
LQ that decays only to $b\nu$, the mass exclusion limit is about $1.1$ TeV, whereas for vector LQs, depending on parameter 
choice, the limits vary from about $1.5$ TeV to about $1.8$ TeV. Another search with two hadronic $\tau$'s and two $b$-jets 
by the CMS collaboration has excluded masses below $1.02$ TeV for a scalar LQ that decays only to $b\tau$ pairs~\cite{CMS:2018eud}. They have 
also performed coupling dependent single production search for such a LQ, that excludes masses below 740 GeV for coupling, 
$\lm^{\ell_q}_{b\tau}=1$~\cite{Sirunyan:2018jdk}. For $\lm^{\ell_q}_{b\tau}>1.4$, this search puts the best limit on the mass 
of such a LQ. Though, strictly speaking, a charge $-1/3$ LQ cannot decay to a $b$-quark and a $\tau$. Hence, the last two bounds 
are not applicable for $\mc S_1$. Some of the limits are also available from the ATLAS searches, but as they are very similar to the CMS 
ones, we do not discuss them here.

The rest of paper is organized as follows. In the next section, we discuss the three scenarios and the new parameters therein. In 
Section \ref{sec:pheno}, we discuss the basic set up for the LHC phenomenology and identify the possible signatures for the three 
scenarios. In Section \ref{relexp}, we discuss the relevant experiments from the LHC and in Section \ref{results} we present our 
main results. Finally, in Section \ref{conclu} we conclude.

\section{The Single Leptoquark Model}\label{sec:mod}

\noindent
The possible interaction terms of $\mc S_1$ that would affect the $R_{D^{(*)}}$ observables can be expressed as follows, 
\begin{eqnarray}\label{eq:Lcompact}
\mc{L} &\supset& \left[\lm_{3\alpha}^L\,\bar{Q}^c_3\left(i\tau_2\right) L_\alpha + \lm_{23}^L\,\bar{Q}^c_2\left(i\tau_2\right) 
L_3 + \lm_{23}^R\,\bar{c}^c\tau_R\right]{\mc S_1^\dagger} \nonumber\\ & & +\ h.c.,
\end{eqnarray} 
where $Q_\alpha(L_\alpha)$ denotes the $\alpha$-th generation quark (lepton) doublet and
$\lm^H_{ab}$ denotes the coupling of $\mc S_1$ with a charge-conjugate quark from generation $a$ and a lepton of 
chirality $H$ from generation $b$. For our analysis, we assume all $\lm$'s to be real without any loss of generality, as 
the LHC data we consider here is insensitive to their complex nature. 

As indicated in the earlier section, we consider two minimal scenarios where the physical state $\mc{S}_1$ is aligned 
either to the up-type quark basis (Scenario-I) or to the down-type quark basis (Scenario-II). From Fig.~\ref{fig:BtoDb}, 
we see that to get a nonzero contribution to the $R_{D^{(*)}}$ observables, we need the couplings of $b\nu \mc S_1$ 
and $c\tau \mc S_1$ interactions to be nonzero. In the two minimal scenarios, these two couplings are not independent -- 
one is generated from the other via the Cabibbo-Kobayashi-Maskawa (CKM) mixing among quarks. As a result, our minimal 
scenarios are completely specified by the LQ mass and just one new coupling. 
For our main analysis, we simply set $\lm^R_{23}=0$ as this coupling alone is not sufficient to address the $R_{D^{(*)}}$ anomalies. 
This is, however, not a bad assumption, since the best-fit values of the corresponding Wilson coefficients do agree with a 
small $\lm^R_{23}$~\cite{Freytsis:2015qca,Cai:2017wry} (presence of a nonzero $\lm^R_{23}$ generates 
new scalar and tensor Wilson operators that are not present in the SM $b\to c\ell\nu_\ell$ transition) and the LHC data is anyway insensitive to 
the $\tau$ polarization. 
Later (in Section \ref{results}) we briefly discuss the limits for nonzero $\lm^R_{23}$ for completeness.
We ignore any mixing in the neutrino sector.
A priori, the large Pontecorvo-Maki-Nakagawa-Sakata (PMNS) mixing in the neutrino sector can generate large interactions 
with the first and second generation neutrinos. Since they all contribute to the missing energy, they are not distinguishable 
at the LHC and hence, their mixing will not affect our analysis. Therefore, we ignore the flavour of neutrinos and 
just denote them as $\nu$.

\subsection{Scenario-I}\noindent 
In this scenario, we assume all the $\lm$'s in Eq.~\eqref{eq:Lcompact} except $\lm_{23}^L$ to be zero. 
Expanding the fermion doublets, we see that this directly generates the $c\tau \mc S_1$ and the $s\nu\mc S_1$ interactions. 
One obtains the $b\nu\mc S_1$ coupling by assuming that in the interaction with $\mc S_1$, the down type quark in $Q_2$ is 
not just the physical $s$-quark, but a mixture of all the down type quarks (i.e., $Q_2$ is in the up-type quark basis). The amount 
of mixing is determined by the CKM matrix elements.  When we move to the mass basis, an effective $b\nu\mc{S}_1$ coupling is generated. 
The effective $b\nu\mc{S}_1$ coupling is CKM suppressed and goes like $\sim V_{cb}\lm_{23}^L$, making the amplitude of the 
process shown in Fig.~\ref{fig:BtoDb} proportional to $\displaystyle V_{cb}\left(\lm_{23}^L\right)^2\approx 4.12\times10^{-2} 
\left(\lm_{23}^L\right)^2$. Though the quark mixing in this case is very similar to that in the SM, there is an important difference. 
Unlike in the SM, here, the larger couplings are off-diagonal in flavour. Written explicitly, the Lagrangian of Eq.~\eqref{eq:Lcompact} 
now looks like,\footnote{The extra couplings generated -- $d\nu \mc S_1$ ($\displaystyle\sim V_{cd} \lm^L_{23}$) and 
$s\nu \mc S_1$ ($\sim V_{cs}\lm^L_{23}$) -- would contribute to known processes with internal LQ interchange(s). For example, 
$B\to (K^{(\ast)},\pi)\nu\bar{\nu}$ and $K\to \pi\nu\bar{\nu}$ 
would receive contributions from LQ~\cite{Deshpande:2004xc,Hiller:2016kry,Cai:2017wry,Angelescu:2018tyl}. However, as mentioned 
in the previous section, we ignore these bounds as our main purpose, in this paper, is to investigate the exclusion limits from the LHC data.} 
\begin{eqnarray}\label{eq:Lsc1}
\mc{L} &\supset& \lm_{23}^L\left[\bar c^c\tau - \lt(V_{cb}\bar b^c + V_{cs}\bar s^c + V_{cd}\bar d^c\rt)\nu\right]{\mc S_1^\dagger}  +\ h.c.\ .\quad 
\end{eqnarray} 
This gives us the following ratio,
\begin{equation}
R_{D^{(*)}}^{\rm I}=\lt |1+C_V^{\rm I}\rt |^2 \times R_{D^{(*)}}^{\rm SM}\,,\label{eq:rds}
\end{equation}
where
\begin{equation}
C_V^{\rm I}=\frac{1}{2\sqrt2 G_{\rm F}V_{cb}}\frac{V_{cb}\left(\lm_{23}^L\right)^2}{2M_{\mc S_1}^2} = \frac{\left(\lm_{23}^L\right)^2}{4\sqrt2 G_{\rm F}M_{\mc S_1}^2} .\label{eq:cv}
\end{equation}
Therefore, one might expect that the favoured values of $\lm_{23}^L$ must be sufficiently large to accommodate the $R_{D^{(*)}}$ 
anomalies, especially for large $M_{\mc S_1}$. This makes it interesting to investigate whether the present LHC data can say something 
about a large $\lm_{23}^L$.

In most of the collider studies of LQs, they are considered to have a generation index, i.e., they are assumed to couple to fermions of 
a specific generation. However, we cannot attach any generation index to $\mc S_1$ in this scenario, since $\mc{S}_1$ couples dominantly 
to a second generation quark and a third generation lepton. From a collider perspective, this leads to an interesting point. A large $\lm_{23}^L$ 
opens up the possibility of producing $\mc S_1$ through $s$- and/or $c$-quark initiated  processes at the LHC. This is a novel aspect in this 
scenario, as, in most of the third generation LQ studies, $b$-quark initiated processes are considered for model dependent productions at the 
LHC~\cite{Sirunyan:2018jdk}, but $b$-PDF (parton distribution function) is much smaller than $s$- or $c$-PDF. This enhances, for example, 
the single production cross section than what is considered in general. It can also give rise to a significant number of $\tau\tau$ or $\tau\nu$ 
events through the $t$-channel $\mc S_1$ exchange processes viz. $cc\to \tau\tau$ or $cs\to \tau\nu$. As a result,  the latest $Z^\prime$ 
resonance search data at the LHC through the $Z^\prime\to \tau\tau$ channel can be used to put bounds on this scenario. Similar bound could 
also be drawn from the $W^\prime\to\tau\nu$ resonance searches.

\subsection{Scenario-II}\noindent
Instead of $\lm_{23}^L$, we now assume that in Eq.~\eqref{eq:Lcompact}, only $\lm^L_{33}$ is nonzero. This directly generates the 
$t\tau \mc S_1$ and $b\nu\mc S_1$ terms. If, like in the previous case, we assume in the $t\tau \mc S_1$ term the top quark is not the 
physical top quark, but a mixture of all the up type quarks (the mixing is once again determined by the CKM matrix) then we obtain an effective 
$c\tau\mc S_1$ coupling of the order of $V_{cb}\lm_{33}^L$ (for simplicity, we ignore the phases in the CKM matrix elements and just consider 
the magnitudes). Now, the ratio of $R_{D^{(*)}}$ would still be given by Eq.~\eqref{eq:rds} but with $\lm_{23}^L\to\lm_{33}^L$ in 
Eq.~\eqref{eq:cv}, i.e.,
\begin{equation}
R_{D^{(*)}}^{\rm II}=\lt |1+C_V^{\rm II}\rt |^2 \times R_{D^{(*)}}^{\rm SM}\,,\label{eq:rds2}
\end{equation}
with
\begin{equation}
C_V^{\rm II}=\frac{1}{2\sqrt2 G_{\rm F}V_{cb}}\frac{V_{cb}\left(\lm_{33}^L\right)^2}{2M_{\mc S_1}^2} = \frac{\left(\lm_{33}^L\right)^2}{4\sqrt2 G_{\rm F}M_{\mc S_1}^2} .\label{eq:cv2}
\end{equation}
Hence, in this scenario too, the new coupling, $\lm^L_{33}$ has to be large to accommodate the  $R_{D^{(*)}}$ anomalies. But, unlike 
before, the $c$-initiated processes would not be large, as it will now come with a suppression by $V_{cb}$. The single production in this 
case would be initiated by the $b$-quark. As a result, the limits from the LHC on the coupling are expected to be weaker than those in the previous case. 
However, in this case, it is possible to identify $\mc S_1$ as a third generation LQ, as it would mainly decay into third generation fermions.

Since we mainly want to study the LHC limits on the couplings relevant for the $R_{D^{(*)}}$ observables, it is now clear that, as far as the LHC phenomenology 
is concerned, Scenario-I has novel features and is more interesting than Scenario-II. In Scenario-I, the limits from the LHC are expected to be on both 
$\lm_{23}^L$ and $M_{\mc{S}_1}$. In contrast, the LHC is mostly insensitive to $\lm_{33}^L$. We have summarized this in Table \ref{tab:scene}. 

\subsection{Scenario-III}\noindent
For completeness, we also consider a next-to-minimal scenario where both $\lm^L_{23}$ and $\lm^L_{33}$ are nonzero. In this case, we can ignore 
the CKM suppressed couplings generated through the quark mixing as both the necessary interactions ($b\nu\mc S_1$ and $c\tau\mc S_1$) for 
explaining the $R_{D^{(*)}}$ 
anomalies are already present. Here, we get the following ratio,
\begin{equation}
R_{D^{(*)}}^{\rm III}=\lt |1+C_V^{\rm III}\rt |^2 \times R_{D^{(*)}}^{\rm SM}\,,\label{eq:rds3}
\end{equation}
with
\begin{equation}
C_V^{\rm III}=\frac{1}{2\sqrt2 G_{\rm F}V_{cb}}\frac{\lm_{23}^L\lm_{33}^L}{2M_{\mc S_1}^2} = \frac{\lm_{23}^L\lm_{33}^L}{4\sqrt2 G_{\rm F}V_{cb}M_{\mc S_1}^2} .\label{eq:cv3}
\end{equation}
Now, of course, none of the $\lm_{23}^L$ and $\lm_{33}^L$ need to be very large to explain the anomalies. Specifically, a moderate $\lm_{23}^L$ (to which the LHC data is sensitive) may be sufficient. 

\begin{table}[t]\vspace{-2.5ex}
\caption{Summary of relevant parameters in various scenarios and sensitivity of the LHC towards them. In Scenario-III, the LHC is indirectly sensitive towards $\lm^L_{33}$ as it can change the total decay width of $\mc S_1$.}\label{tab:scene}
\begin{tabular}{| c | c | c |}\hline
Scenario &	Parameters &	LHC Sensitivity\\\hline\hline
I& $M_{\mc S_1}$, $\lm^L_{23}$  & $M_{\mc S_1}$, $\lm^L_{23}$\\
II& $M_{\mc S_1}$, $\lm^L_{33}$  & $M_{\mc S_1}$\\
III& $M_{\mc S_1}$, $\lm^L_{23}$, $\lm^L_{33}$  & $M_{\mc S_1}$, $\lm^L_{23}$, $\lt(\lm^L_{33}\rt)$\\\hline
\end{tabular}
\end{table}

\section{LHC Phenomenology: The Preliminaries}
\label{sec:pheno}

\noindent
To study the LHC signatures of the three scenarios, we make use of various publicly available packages.
We first implement the new terms in the Lagrangian in \textsc{FeynRules}~\cite{Alloul:2013bka} to create the Universal FeynRules Output (UFO)~\cite{Degrande:2011ua} model files suitable for \textsc{MadGraph5}~\cite{Alwall:2014hca}. In \textsc{MadGraph5}, we use the NNPDF2.3LO~\cite{Ball:2012cx} PDF set to generate all the signal and the background events. For signal events we set the factorization scale and the renormalization scale, $\mu_{\rm F} = \mu_{\rm R} = M_{\mc{S}_1}$. 
The scales are kept fixed at the highest scale for each background  process. Subsequent parton showering and hadronization of the events are done using \textsc{Pythia6}~\cite{Sjostrand:2006za}. The detector environment effects are simulated with \textsc{Delphes}3 \cite{deFavereau:2013fsa}. The jets are clustered using the anti-$k_{\rm T}$ algorithm \cite{Cacciari:2008gp} with 
radius $R = 0.4$ with the help of the \textsc{FastJet} \cite{Cacciari:2011ma} package within \textsc{Delphes}3. For our analysis, all the event samples are generated at the leading order. However, we multiply the pair production cross sections by a typical next-to-leading order (NLO) QCD 
$K$-factor of $1.3$ (as available in the literature, see, e.g., Ref.~\cite{Mandal:2015lca}). 

In the minimal scenarios, all the CKM suppressed effective couplings that are generated by the quark
mixing play a negligible role at the LHC. 
In Scenario-I, $\mc{S}_1$ dominantly decays to $c\tau$, $s\nu$ final states via $\lm_{23}^L$ with about 50\% branching fraction in each mode, producing yet unexplored signatures at the LHC. The pair production of $\mc{S}_1$ leads to the following final states:
\begin{equation}
\left.
\begin{array}{rcccl}
\mc{S}_1\mc{S}_1 &\to& c\tau\ubr{-2.5}\ c\tau\ubr{-2.5}&\equiv& \tau\tau+2j \\
\mc{S}_1\mc{S}_1 &\to& c\tau\ubr{-2.5}\ s\nu\ubr{-2.5}&\equiv& \tau+2j+\slashed{E}_{\rm T} \\
\mc{S}_1\mc{S}_1 &\to& s\nu\ubr{-2.5}\ s\nu\ubr{-2.5}&\equiv& 2j+\slashed{E}_{\rm T}
\end{array}\right\},
\end{equation}
where the curved connection above a pair of particles indicates that they are coming from the decay of 
an $\mc{S}_1$ and $j$ denotes a light jet. In addition to the pair production, there are other production channels like the single ($pp\to\mc S_1\tau$, $\mc S_1\nu$ etc.) and the indirect productions of $\mc{S}_1$ ($pp\to\tau\tau$, $\tau\nu$ or $\nu\nu$ through the $t$-channel $\mc{S}_1$ exchange, mainly $s$- and/or $c$-quark initiated) that could have detectable signatures at the LHC. All these processes have very different kinematics, but, if we look only at the final state signatures, all of them would have some or all of the following three kinds:
\begin{align*}
(i)~\tau\tau+jets,\quad(ii)~\tau+\slashed{E}_{\rm T}+jets\quad{\rm and}\quad (iii)~\slashed{E}_{\rm T}+jets.
\end{align*}
Here, ``$jets$'' stands for any number ($\geq0$) of untagged jets (including $b$-jets that are not tagged). Among these, due to the absence of any identifiable charged lepton in the final state, the bounds from the $\slashed{E}_{\rm T}+jets$ channel are expected to be weaker than those obtained from the other two (this signature has been considered before in Ref.~\cite{Biswas:2018snp}, albeit for a different LQ species). Hence, in this paper, we focus only on the first two signatures, i.e., $\tau\tau+jets$ and $\tau+\slashed{E}_{\rm T}+jets$. 

\begin{figure}[t]\vspace{-1.5ex}
\includegraphics[width=0.48\textwidth]{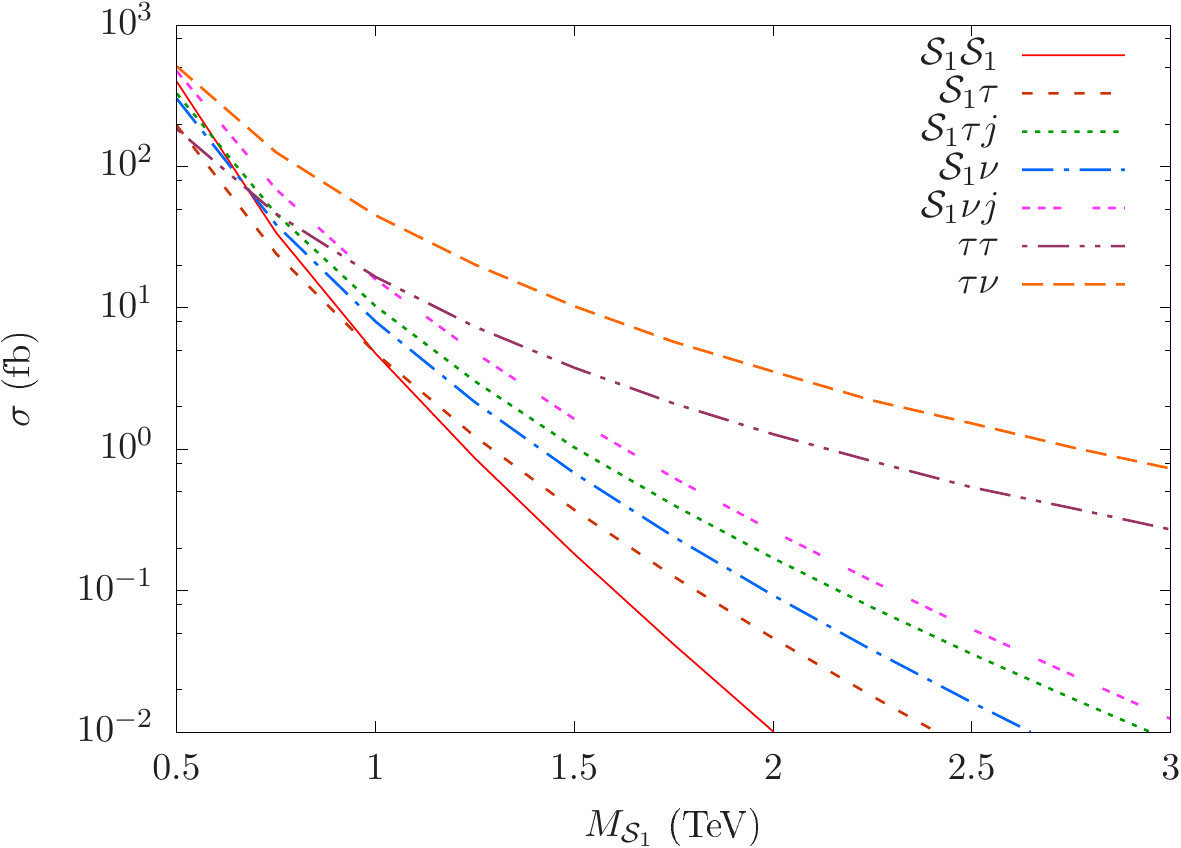}
\caption{The parton-level cross sections of different production channels of $\mc{S}_1$  at the 13 TeV LHC as functions of
$M_{\mc{S}_1}$ in Scenario-I. These cross sections are computed for a benchmark coupling
$\lm_{23}^L=1$. Here, the $j$ in the $\mc S_1\tau j$ and $\mc S_1\nu j$ processes includes all the light jets as well as $b$ jets. Their cross sections are generated with a cut on the transverse momentum of the jet, $p^j_{\rm T}>20$ GeV.}
\label{fig:cs}
\end{figure}

The single productions that contribute to the $\tau\tau+jets$ or the $\tau+\slashed{E}_{\rm T}+jets$ final states have two different topologies~\cite{Mandal:2015vfa} -- (a) Born single (BS, where an $\mc{S}_1$ is produced in association with a lepton, i.e., $\mc{S}_1\tau$ or $\mc S_1\nu$) and (b) new subprocesses of three-body single production (NS3, where there is an extra (hard) jet in addition to the lepton, i.e., $\mc{S}_1\tau j$ and $\mc{S}_1\nu j$). The lepton-jet pair in the $\mc{S}_1\tau j$ or $\mc{S}_1\nu j$  final states might come from the decay of another $\mc S_1$ (in which case, the process is essentially the pair production) or the jet could be an initial or final state radiation (ISR/FSR) emitted from a BS process; but we do not count them in NS3, 
only subprocesses with completely new topologies are considered. For our main analysis, we compute the contributions of inclusive single productions. For a specific final state, this
includes the combined contributions of all the single productions contributing to that final state. While computing the inclusive signals, we define NS3 in this manner to avoid double counting BS+ISR/FSR contribution in the three body single production again. A detailed discussion on how one can systematically estimate the inclusive single production cross sections is presented in Ref.~\cite{Mandal:2015vfa}. We deploy the technique of the matrix element-parton shower matching (ME$\oplus$PS) to estimate them. More specifically, we combine the following processes using the MLM matching technique~\cite{Mangano:2006rw},
\begin{eqnarray}
&&\underline{\tau\tau+jets:}\nn\\
&&\left. \begin{array}{lclcl}
pp &\to &\mc{S}_1\ \tau &\to & \tau j\ubr{-2.5}\ \tau \\
pp &\to &\mc{S}_1\ \tau j &\to & \tau j\ubr{-2.5}\ \tau j \label{eq:match_tautau}\\
pp &\to &\mc{S}_1\ \tau jj &\to & \tau j\ubr{-2.5}\ \tau jj
\end{array}\right\}, \\
&&\underline{\tau+\slashed{E}_{\rm T}+jets:}\nn\\
&&\left. \begin{array}{lcccl}
pp &\to &(\mc{S}_1\ \tau + \mc{S}_1\ \nu) &\to & \nu j\ubr{-2.5}\ \tau + \tau j\ubr{-2.5}\ \nu\\
pp &\to &(\mc{S}_1\ \tau j+ \mc{S}_1\ \nu j) &\to & \nu j\ubr{-2.5}\ \tau j + \tau j\ubr{-2.5}\ \nu j\label{eq:match_taunu}\\
pp &\to &\hspace{-0.75ex}(\mc{S}_1\ \tau jj+ \mc{S}_1\ \nu jj) \hspace{-0.75ex}&\to & \nu j\ubr{-2.5}\ \tau jj + \tau j\ubr{-2.5}\ \nu jj
\end{array}\right\},\quad\ 
\end{eqnarray}
with a matching scale $Q_{cut}\sim 50$ GeV for all LQ masses.

Let us now have a look at the strengths of the different production modes of $\mc{S}_1$ at the LHC. In Fig.~\ref{fig:cs}, we show the parton-level cross sections of different production channels (viz. the
pair, the single and the indirect productions) of $\mc{S}_1$ in Scenario-I at the 13 TeV LHC for a varying $M_{\mc{S}_1}$. The pair production cross section is large for $M_{\mc{S}_1}\lesssim1$ TeV because of the large gluon PDF in the small-$x$ region. On the other hand, it is well known that the single and the indirect productions dominate over the pair production for heavier LQ masses due to less phase space suppression (see~\cite{Mandal:2012rx,Mandal:2015vfa}). 
Of course, the cross sections of these processes also depend on the strength of the new physics couplings. 
For ease of notation, in the rest of the paper, we shall denote the free couplings of the minimal scenarios as $\lm$. It is equal to $\lm_{23}^L$ in Scenario-I and $\lm_{33}^L$ in Scenario-II. Since, the LHC is rather insensitive to $\lm_{33}^L$, in Scenario-III also, we refer to $\lm^L_{23}$ as $\lm$. 

In Fig.~\ref{fig:cs}, we use a benchmark coupling, $\lm=1$. The pair production ($pp\to\mc{S}_1\mc{S}_1$) is mostly governed by the strong $\mathrm{SU}(3)_c$ coupling, and is almost insensitive to small 
$\lm$ (the $\lm$ dependence comes through the $t$-channel lepton exchange diagram). However, since the overall pair production contribution in our results is small compared to other processes, we shall ignore this $\lm$ dependent piece in the rest of the paper. For the single productions as well as the indirect productions, the leading order dependence on $\lm$ is easily factorizable. The single production cross sections are proportional to $\lm^2$. Just like the process in Fig.~\ref{fig:BtoDb}, the amplitudes of the indirect production processes, $pp\to\tau\tau,$ $\tau\nu$ through the $t$-channel 
$\mc{S}_1$ exchange are proportional to $\lm^2$ leading to a $\lm^4$ contribution to the cross section. Interestingly, the $t$-channel LQ exchange processes interfere with the exclusive $pp\to\tau\tau, \tau\nu$ processes in the SM (mediated by the $s$-channel exchange of electroweak vector bosons at the leading order) as they share same initial and final states. The interference is destructive and is of $\mc O(\lm^2)$. Hence, in Scenario-I, the total exclusive $pp\to\tau\tau/\tau\nu$ cross section can be expressed as,
\begin{equation}
\sg^{excl}_{pp\to xy} = \sg_{\rm SM} - \lm^2\ \sg_{\times}\left(M_{\mc S_1}\right) + \lm^4\ \sg_{t}\left(M_{\mc S_1}\right)\label{eq:intf}
\end{equation}
where $xy \in \{\tau\tau, \tau\nu\}$, $\sg_{\times}$ and $\sg_{t}$ are the interference and the pure $t$-channel BSM contribution to $\sg^{excl}_{pp\to xy}$ at $\lm=1$, respectively. The minus sign of the interference contribution $\sg_{\times}$ indicates its destructive nature. Both of these terms are functions of $M_{\mc S_1}$. In  Fig.~\ref{fig:cs}, we only show the $\sg_{ t}$ part for $pp\to\tau\tau$ or $\tau\nu$. 

Roughly speaking, we are interested in the parameter space where $M_{\mc S_1} \gtrsim 1$ TeV and $\lm\gtrsim 1$ (as, na{\"i}vely, different direct LQ searches seem to suggest that a LQ with mass in the sub-TeV regime is less likely to exist and the $R_{D^{(*)}}$ anomalies hint towards a large $\lm$). It is clear from Fig.~\ref{fig:cs} that in this region, the single and the indirect productions are more important than the pair production. This figure, however, does not give the full picture -- there remain two more important points to consider before recasting the experimental bounds. 
\begin{enumerate}[wide, labelwidth=!, labelindent=0pt]
\item
Since different production modes have different kinematics, in any experiment the selection efficiencies (the fraction of events that survives the selection criteria) in these modes would be different. Hence, once the kinematic cuts are applied, the ratios among number of events passing through the cuts are, in general, different from the corresponding ratios of cross sections.
\item
The interference contributions depend on the size of the SM contribution to the $pp\to\tau\tau$ or $\tau\nu$ processes. It is normally much larger than the new (purely BSM) contributions. Hence, in parts of the parameter space, it is possible that the interference term dominates over all other modes and
contributions (i.e., $\lm^2 \sg_{\times} > \lm^4 \sg_{\rm t},\lm^2\sg_{s}^{incl},\sg_{p}$ where 
$\sg^{incl}_s$ is the inclusive single production cross section at $\lm=1$ and $\sg_p$ is the pair production cross section). It would then lead to a reduction in the expected number of events in the $\tau\tau+jets$ or $\tau +\slashed{E}_{\rm T}+jets$ channels than the SM only case. This, of course, depends also on the model/scenario as well as the part of phase space we are looking at. As we shall see later, this will
happen in Scenario-I, but, in Scenario-II, where $\mc{S}_1$ dominantly couples with the third generation
quarks such a situation would not arise.\footnote{The fact that in the dilepton or the monolepton channels some species of LQs can significantly interfere (constructively or destructively) with the SM background is known~\cite{Wise:2014oea,Raj:2016aky,Bansal:2018eha}. In particular, Ref.~\cite{Bansal:2018eha} has recently used the interference spectra in the charged-current Drell-Yan (monolepton) channel to obtain the projected bounds on the LQs that couples with electrons for the future high luminosity LHC runs.}
\end{enumerate}

In Scenario-II, the dominant signatures of the pair production will be the following,
\begin{equation}\left.
\begin{array}{rcccl}
\mc{S}_1\mc{S}_1 &\to& t\tau\ubr{-2.5}\ t\tau\ubr{-2.5}&\equiv& tt+\tau\tau \\
\mc{S}_1\mc{S}_1 &\to& t\tau\ubr{-2.5}\ b\nu\ubr{-2.5}&\equiv& t+\tau+j+\slashed{E}_{\rm T} \\
\mc{S}_1\mc{S}_1 &\to& b\nu\ubr{-2.5}\ b\nu\ubr{-2.5}&\equiv& 2j+\slashed{E}_{\rm T}
\end{array}\right\}.
\end{equation}
However, unlike Scenario-I, in this case, the single and the indirect productions would be suppressed because of the smallness of $b$-PDF in the initial states. Hence, we do not discuss the signatures of these productions modes for this scenario. As already indicated (see Table \ref{tab:scene}), in this case, the only significant bound from the current LHC data would be on $M_{\mc{S}_1}$, not on the coupling $\lm$.

In Scenario-III, all the processes mentioned for Scenario-I and II would be present. As long as the coupling $\lm^L_{23}$ is not small, the total contribution to the $\tau\tau+jets$ or $\tau+\slashed E_{\rm T}+jets$ final states would be significant.

\section{Relevant Experiments at the LHC}
\label{relexp}

\noindent
Since in Scenario-I, all the production processes contribute to the $\tau\tau+jets$ and the $\tau+\slashed E_{\rm T}+jets$ final states, we consider the latest $pp\to Z'\to\tau\tau$ +and $pp\to W'\to \tau\nu$ searches at the LHC~\cite{Aaboud:2017sjh,Aaboud:2018vgh,Khachatryan:2016qkc,Sirunyan:2018lbg} to constrain the LQ parameters. We notice that these searches do not put any restriction on the number jets and just look for the $\tau\tau+jets$ or the $\tau+\slashed{E}_{\rm T}+jets$ signatures -- exactly as we want.  Below, we review the essential details of the ATLAS searches~\cite{Aaboud:2017sjh,Aaboud:2018vgh} (since, the ATLAS and the CMS searches are similar, we consider the ATLAS searches only).\\

\noindent
\underline{{\bf ATLAS $\tau\tau$ search}~\cite{Aaboud:2017sjh}}: A search for heavy resonance in the $\tau\tau$ channel was performed by the ATLAS collaboration at the 13 TeV LHC
with $36$ fb$^{-1}$ integrated luminosity. In this analysis, events are categorized on the basis of $\tau$-decays: $\tau_{had}\tau_{had}$ mode where both the $\tau$'s decay hadronically
and $\tau_{lep}\tau_{had}$ mode where one $\tau$ decays leptonically and the other one decays hadronically. Following Ref.~\cite{Aaboud:2017sjh}, 
we outline the \emph{ basic event selection criteria for the $\tau\tau+jets$ channel} that we shall also use in our analysis:
\begin{itemize}[wide, labelwidth=!, labelindent=0pt]
\item 
{In the $\tau_{had}\tau_{had}$ channel, there must be
\begin{itemize}
\item at least two hadronically decaying $\tau$'s are tagged with no electrons or muons,
\item two $\tau_{had}$'s have $p_{\rm T}(\tau_{had})>65$ GeV, they are oppositely charged and separated in the azimuthal plane by $|\Delta\phi(p_{\rm T}^{\tau_1},p_{\rm T}^{\tau_2})|>2.7$
rad.
\end{itemize}
}

\item
In the $\tau_{lep}\tau_{had}$ channel, in addition to one $\tau_{had}$, any event must contain only one $\ell=e,\mu$ such that
\begin{itemize}
\item the hadronic $\tau_{had}$ must have $p_{\rm T}(\tau_{had})>25$ GeV and $|\eta(\tau_{had})|<2.3$ (excluding $1.37 < |\eta| < 1.52$),
\item
if the lepton is an electron then $|\eta| < 2.4$ (excluding $1.37 < |\eta| < 1.52$) and if it is a muon then $\eta < 2.5$,
\item the lepton must have $p_{\rm T}(\ell)>30$ GeV and its azimuthal separation from the $\tau_{had}$ must be $|\Delta\phi(p_{\rm T}^{\ell},p_{\rm T}^{\tau_{had}})|>2.4$ rad.
\item A cut on the transverse mass, $m_{\rm T}(p_{\rm T}^\ell, \slashed E_{\rm T})>40$ GeV of the selected lepton and the missing transverse momentum is applied, where transverse mass is defined as,
\begin{equation}
m_{\rm T}(p_{\rm T}^A, p_{\rm T}^B) = \left[2p_{\rm T}^A p_{\rm T}^B\lt\{1- \cos\Delta\phi(p_{\rm T}^A, p_{\rm T}^B)\rt\}\right]^{1/2}.
\end{equation}
\end{itemize}

\end{itemize}
In the analysis, another quantity, the total transverse mass, is also defined,
\begin{eqnarray}
m^{\rm tot}_{\rm T}\left(\tau_1,\tau_2,\slashed E_{\rm T}\right) &=& \left[m^2_{\rm T}(p_{\rm T}^{\tau_1}, p_{\rm T}^{\tau_2})+m^2_{\rm T}(p_{\rm T}^{\tau_1}, \slashed E_{\rm T})\right.\nonumber\\&&\ +\left.m^2_{\rm T}(p_{\rm T}^{\tau_2},\slashed E_{\rm T})\right]^{1/2}, 
\end{eqnarray}
where $\tau_2$ in the $\tau_{lep}\tau_{had}$ channel represents the lepton.
A distribution of the observed and the SM events with respect to $m^{\rm tot}_{\rm T}$ is presented in the analysis. \\

\noindent
\underline{{\bf ATLAS $\tau\nu$ Search}~\cite{Aaboud:2018vgh}}: 
The ATLAS search in the $\tau\nu$ channel has been performed with $36.1$ fb$^{-1}$ integrated luminosity at the 13 TeV LHC. Only hadronically decaying $\tau$ leptons ($\tau_{had}$) are considered for the analysis. Below, we show the \emph{basic event selection criteria for the $\tau+\slashed E_{\rm T}+jets$ channel}:
\begin{itemize}[wide, labelwidth=!, labelindent=0pt]

\item At least one $\tau_{had}$ with transverse momentum $p_{\rm T}(\tau_{had})>50$ GeV and $|\eta(\tau_{had})|<2.4$ is required.

\item Any event must have missing transverse energy, $\slashed{E}_{\rm T}>150$ GeV with $\displaystyle 0.7 < p_{\rm T}(\tau_{had})/\slashed{E}_{\rm T} < 1.3$.

\item The azimuthal angle between $\vec{p}_{\rm T}(\tau_{had})$ and $\vec{\slashed{E}}_{\rm T}$ i.e. 
$\Dl\phi\lt(\vec{p}_{\rm T}(\tau_{had}),\vec{\slashed{E}}_{\rm T}\rt)> 2.4$.

\item Events are rejected if they contain any electron or muon with 
$p_{\rm T}(e) > 20$ GeV, $|\eta(e)| < 2.47$ (excluding the barrel-endcap region, $1.37 < |\eta(e)| < 1.52$) or $p_{\rm T}(\mu) > 20$ GeV, $|\eta(\mu)| < 2.5$.

\end{itemize}
In this analysis, the distribution of the events with respect to a varying transverse mass, $m_{\rm T}\lt(\vec{p}_{\rm T}(\tau_{had}),\vec{\slashed{E}}_{\rm T}\rt)$ is given. \\

In addition, for Scenario-I, we also take into account the CMS search in the $2j+\slashed{E}_{\rm T}$ channel~\cite{Sirunyan:2018kzh} assuming the jets in this case are originating from $s$ quarks (pair production).  
For Scenario-II, we recast the CMS searches for the pair production of third generation LQ with $tt\tau\tau$~\cite{Sirunyan:2018nkj} and $bb+\slashed{E}_{\rm T}$~\cite{Sirunyan:2018kzh} final states. For Scenario-III, we use all these experiments together to recast the exclusion limits in the $\lm^L_{23}-M_{\mc S_1}$ plane for fixed $\lm_{33}^L$.\\

Before we move on to the actual recast we quickly take note of some other related experiments.

\begin{enumerate}[wide, labelwidth=!, labelindent=0pt]
\item In principle, the searches for a heavy charged gauge boson ($W'$) together with a heavy neutrino ($N$) through the process $pp\to W'\to \tau N\to \tau\tau jj$~\cite{Sirunyan:2018vhk} could also be considered like the $pp\to Z^\prime\to\tau\tau$ process for Scenario-I. However, since these searches explicitly look for two hard jets in the final states, they disfavour all production modes except the pair production which is anyway small compared to the others. 
 
\item The searches for the pair production of a third generation charge $2/3$ LQ in the $\tau\tau bb$ final state by CMS~\cite{Khachatryan:2016jqo,Sirunyan:2017yrk,CMS:2018eud} (or even its single production in the $\tau\tau b$~\cite{Sirunyan:2018jdk} channel) cannot be used easily for recasting, since that would require relaxing the explicit requirement of $b$-tagging in the final state jet(s) (i.e., treating the final states as $\tau\tau jj$ or $\tau\tau j$). 
\end{enumerate}

\begin{table*}[t]\vspace{-1.5ex}
\caption{\label{tab:tautau} Cross sections ($\sg$) in fb of various production processes of $\mc{S}_1$ that contribute to the $\tau\tau+jets$ channel at the $13$ TeV LHC for $M_{\mc{S}_1}=1$, $1.5$, $2$ TeV with $\lm=1$ (except the pair production) in Scenario-I. The single and pair production cross sections include appropriate branching fractions. To reduce the contamination from the $Z$-peak, the indirect production events are generated with an invariant mass cut,  $M(\tau,\tau)>140$ GeV, applied at the generator level. The negative signs in front of $\sg_{\times}$ and $N_{\times}$ indicate destructive interference (see text). The efficiencies ($\varepsilon$, shown in percentage)  are obtained after applying the cuts~\cite{Aaboud:2017sjh} listed in the \emph{basic event selection criteria for the $\tau\tau+jets$ channel} in Section-\ref{relexp}. We scale the number of events with the efficiencies to obtain the number of events passing through the cuts  for the experimental luminosity, $\mc{L}=36$ fb$^{-1}$ as $N_i=\mc L\sg_i\,\varepsilon_i$. The NLO cross sections for the pair production are obtained with a typical $K$-factor of $1.3$. In the inclusive single production, $\sg_{nj}$ with $n=0$, $1$, $2$ are the  parton-level cross sections of the processes defined in Eq.~\eqref{eq:match_tautau} generated with a low $p_{\rm T}$ cut on the jets. The matched inclusive cross section of $pp\to\mc S_1\tau\to \tau\tau j$ process is shown as $\sg_{s,\tau}^{incl}$ (see Section-\ref{sec:pheno}).}
\begin{center}
\begin{tabular}{|c|c|c|c|c|c|c|c|c|c|c|c|c|c|c|c|}
\hline
\multirow{3}{*}{$M_{\mc{S}_1}$} & \multicolumn{3}{c|}{Pair (NLO)} & \multicolumn{6}{c|}{Indirect (fiducial)} & \multicolumn{6}{c|}{Inclusive single} \\ \cline{2-16} 
& \multicolumn{3}{c|}{$(\lm\approx 0)$} & \multicolumn{3}{c|}{Interference ($\lm^2,\lm=1$)} & \multicolumn{3}{c|}{BSM ($\lm^4,\lm=1$)} & \multicolumn{6}{c|}{$\mc{S}_1\tau$ ($\lm^2,\lm=1$)}\\ \cline{2-16} 
(TeV) & $\sg_p$  & \hspace{0.2cm}$\varepsilon_p$\hspace{0.2cm}  & $N_p$ & $-\sg_{\times}$  & \hspace{0.2cm}$\varepsilon_{\times}$\hspace{0.2cm} & $-N_{\times}$ & $\sg_t$ & $\varepsilon_t$  & $N_t$ & $\sg_{0j}$  & $\sg_{1j}$  & $\sg_{2j}$  & $\sg_{s,\tau}^{incl}$  & $\varepsilon_{s,\tau}^{incl}$ & $N_{s,\tau}^{incl}$\\ \hline
$1.0$ & $1.329$ & $3.4$ & $1.63$ & $-58.37$ & $5.6$ & $-117.7$ & $7.941$ & $9.7$ & $27.73$ &$2.627$ & $5.218$ & $2.923$ & $6.679$ & $3.6$ & $8.66$  \\ \hline
$1.5$ & $0.052$ & $3.2$ & $0.06$ & $-26.86$ & $5.7$ & $-55.12$ & $1.880$ & $10.2$ & $6.90$ & $0.225$ & $0.674$ & $0.364$ & $0.734$ & $3.3$ & $0.87$\\ \hline
$2.0$ & $0.003$ & $3.1$ & $0.00$ & $-15.30$ & $5.7$ & $-31.40$ & $0.651$ & $10.0$ & $2.34$ & $0.031$ & $0.139$ & $0.073$ & $0.134$ & $3.3$ & $0.16$\\ \hline
\end{tabular}
\end{center}\end{table*}

\begin{table*}[t]\vspace{-1.5ex}
\caption{\label{tab:taunu} Cross sections ($\sg$) in fb of various production processes of $\mc{S}_1$ that contribute to the $\tau+\slashed E_{\rm T}+jets$ channel at the $13$ TeV LHC for $M_{\mc{S}_1}=1$, $1.5$, $2$ TeV with $\lm=1$ (except the pair production) in Scenario-I. The single and pair production cross sections include appropriate branching fractions. To reduce the contamination from the $W$-peak, the indirect production events are generated with a  $m_{\rm T}(\tau,\nu)>250$ GeV cut applied at the generator level. The efficiencies  are obtained after applying the cuts~\cite{Aaboud:2018vgh} listed in the \emph{basic event selection criteria for the $\tau+\slashed E_{\rm T}+jets$ channel} in Section-\ref{relexp}. With these we obtain the number of events passing through the cuts for $36.1$ fb$^{-1}$ of luminosity.
The NLO cross sections for the pair production are obtained with a typical $K$-factor of $1.3$. In the inclusive single production, $\sg_{nj}$ with $n=0$, $1$, $2$ are the  parton-level cross sections of the processes defined in Eq.~\eqref{eq:match_taunu} generated with a low $p_{\rm T}$ cut on the jets. The matched inclusive cross section of $pp\to\mc S_1\tau+\mc S_1\nu\to \tau\nu j$ process is shown as $\sg_{s,\tau}^{incl}$ (see Section-\ref{sec:pheno}).}
\begin{tabular}{|c|c|c|c|c|c|c|c|c|c|}
\hline
\multirow{3}{*}{$M_{\mc{S}_1}$} & \multicolumn{3}{c|}{Pair (NLO)} & \multicolumn{6}{c|}{Indirect (fiducial)} \\ \cline{2-10} 
& \multicolumn{3}{c|}{($\lm\approx 0$)} & \multicolumn{3}{c|}{Interference ($\lm^2,\lm = 1$)} & \multicolumn{3}{c|}{BSM ($\lm^4,\lm=1$)} \\ \cline{2-10} 
(TeV) & $\sg_p$ & \hspace{0.2cm}$\varepsilon_p$\hspace{0.2cm}  & $N_p$ & $-\sg_{\times}$ & \hspace{0.2cm}$\varepsilon_{\times}$\hspace{0.2cm} & $-N_{\times}$ & $\sg_t$ & \hspace{0.2cm}$\varepsilon_t$\hspace{0.2cm} & $N_t$ \\ \hline
$1.0$ & $2.662$ & $1.5$ & $1.44$ & $-54.44$ & $4.0$ & $-78.61$ & $13.38$ & $7.6$ & $36.71$ \\ \hline
$1.5$ & $0.104$ & $1.6$ & $0.06$ & $-27.57$ & $4.1$ & $-40.81$ & $3.469$ & $8.7$ & $10.90$ \\ \hline
$2.0$ & $0.007$ & $1.5$ & $0.00$ & $-16.00$ & $4.3$ & $-24.84$ & $1.238$ & $8.7$ &  $3.89$ \\ \hline
\end{tabular}
\begin{tabular}{|c|c|c|c|c|c|c|c|c|c|c|c|c|}
\hline
\multirow{3}{*}{$M_{\mc{S}_1}$} & \multicolumn{12}{c|}{Inclusive single} \\ \cline{2-13} 
& \multicolumn{6}{c|}{$\mc{S}_1\tau\to\tau\nu j$ ($\lm^2,\lm=1$)} & \multicolumn{6}{c|}{$\mc{S}_1\nu\to\tau\nu j$ ($\lm^2,\lm=1$)} \\ \cline{2-13} 
(TeV) & $\sg_{0j}$ & $\sg_{1j}$ & $\sg_{2j}$  & $\sg_{s,\tau}^{incl}$ & $\varepsilon_{s,\tau}^{incl}$ & $N_{s,\tau}^{incl}$ & $\sg_{0j}$  & $\sg_{1j}$ & $\sg_{2j}$  & $\sg_{s,\nu}^{incl}$  & $\varepsilon_{s,\nu}^{incl}$ & $N_{s,\nu}^{incl}$ \\ \hline
$1.0$ & $2.634$ & $5.212$ & $2.906$ & $6.651$ & $1.4$ & $3.36$ & $5.448$ & $10.25$ & $4.770$ & $12.52$ & $0.9$ & $4.07$ \\ \hline
$1.5$ & $0.226$ & $0.674$ & $0.366$ & $0.732$ & $1.4$ & $0.37$ & $0.682$ & $1.803$ & $0.785$ & $1.822$ & $0.9$ & $0.59$ \\ \hline
$2.0$ & $0.031$ & $0.139$ & $0.072$ & $0.132$ & $1.3$ & $0.06$ & $0.156$ & $0.537$ & $0.223$ & $0.471$ & $0.9$ & $0.15$ \\ \hline
\end{tabular}

\end{table*}

\begin{figure*}[t]\vspace{-1.5ex}
\subfloat[]{\includegraphics[width=0.48\textwidth]{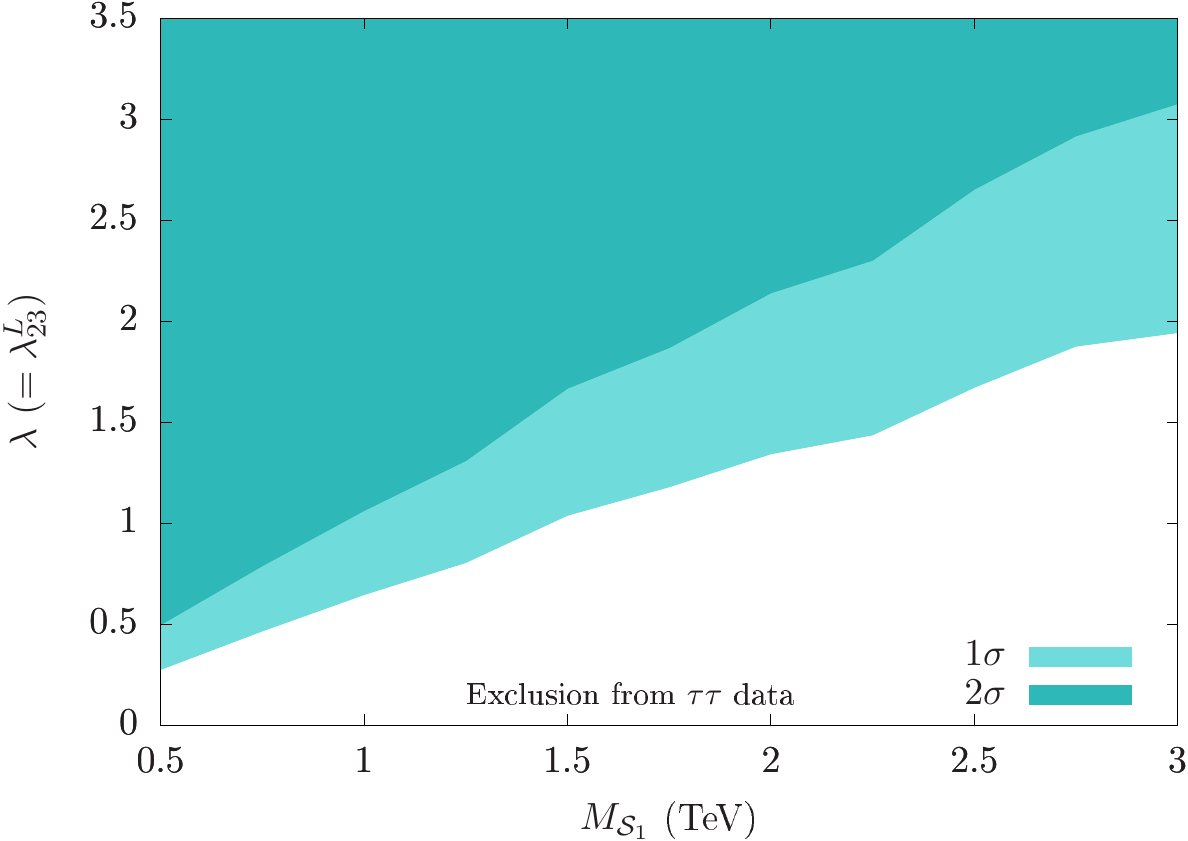}\label{fig:lmexclutautau}}\hfill
\subfloat[]{\includegraphics[width=0.48\textwidth]{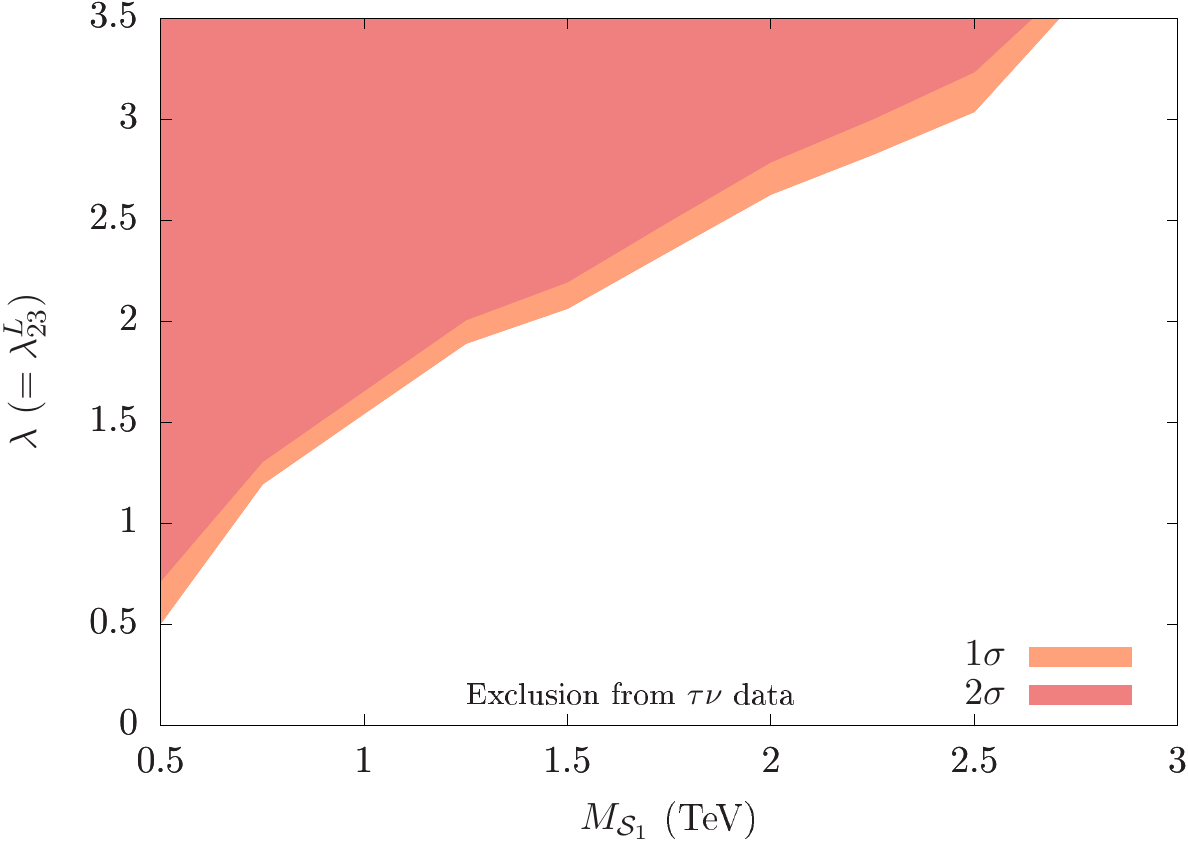}\label{fig:lmexclutaunu}}
\caption{The $1\sg$ and $2\sg$ CL exclusion limits on $\lm=\lm^L_{23}$ in Scenario-I as functions of $M_{\mc{S}_1}$ using the ATLAS
(a) $\tau\tau$~\cite{Aaboud:2017sjh}  and (b) $\tau\nu$~\cite{Aaboud:2018vgh} resonance search data. The coloured regions are excluded. We keep $\lm\leq3.5$ to ensure $\displaystyle\lm^2/4\pi< 1$.}
\label{fig:lmexclu}
\end{figure*}
\section{Data Recast and Exclusion Limits} 
\label{results}

\noindent
We have validated all of our analysis codes by reproducing some relevant simulation results from both the ATLAS and the CMS analyses. We have estimated the cut efficiencies ($\varepsilon$, fraction of events surviving the cuts) of these channels by mimicking the cuts used in the ATLAS searches. 
As we process the events through the detector simulator before computing $\varepsilon$, it has to be compared with the $\textrm{acceptance}\times\textrm{efficiencies}$ presented in the experimental analyses to be precise. However, we will refer to it loosely as the efficiency in this paper.  
We have generated $pp\to Z^\prime (W^\prime)\to \tau\tau(\tau\nu)$ (for the sequential $Z^\prime(W^\prime)$ model) events for some benchmark $Z^\prime(W^\prime)$ masses. We find that the cut-efficiencies we obtain with these are in close agreement with those in Refs.~\cite{Aaboud:2017sjh,Aaboud:2018vgh,Aad:2015osa,Sirunyan:2018lbg}.\footnote{We quote a random example to demonstrate the agreement. For 35.9 fb$^{-1}$ of integrated luminosity, we find $10619$ events (generated for a benchmark mass $M_{W^\prime}=1$ TeV) passing through the selection cuts of Ref.~\cite{Sirunyan:2018lbg} that are in the range $0.5~\textrm{TeV}<m_{\rm T}<1~\textrm{TeV}$. This is to be compared to $10079\pm 1581$ (simulated) events as reported in Ref.~\cite{Sirunyan:2018lbg}.}

As observed in the last section, the experimentally observed (total) transverse mass distributions are available for these channels along with the bin-wise SM only contributions from the two ATLAS searches~\cite{Aaboud:2017sjh,Aaboud:2018vgh}. We use these distributions to estimate the experimental limits on $\lm^L_{23}$ in Scenario-I \& III. For that, first, we apply the basic selection cuts to our simulated signal events (i.e., events from the various production channels mentioned before) in these scenarios 
for both the $\tau\tau+jets$ and $\tau+\slashed{E}_{\rm T}+jets$ channels.  

\subsection{Bounds on Scenario-I}
\noindent
We show the cross sections of various production channels of $\mc{S}_1$  for $\lm=\lm^L_{23}=1$ and $\lm^L_{33}=0$ (i.e., for Scenario-I), the corresponding efficiencies and the number of events surviving the cuts in Tables~\ref{tab:tautau} and \ref{tab:taunu}, respectively. The negative signs in the interference cross sections ($-\sg_{\times}$) signify its destructive nature. We see that the contributions of the inclusive single productions [Eqs.~\eqref{eq:match_tautau} and \eqref{eq:match_taunu}] are small but non-negligible. Hence, one cannot completely ignore them 
while setting limits on $\lm$. The pair productions are $\lm$-insensitive and their contributions are negligible. 
There are a few  points to note here. 
\begin{enumerate}[wide, labelwidth=!, labelindent=0pt]
\item 
The selection cuts used in the experimental analyses we are considering are optimized for an $s$-channel resonance. In our case, all the production processes including the $t$-channel $\mc S_1$ exchange have a different topology. Hence, the cut-efficiencies becomes relatively smaller. 
We can see that the number of surviving events after the cuts, the contribution of the indirect production is largest among all the production processes.
\item
In the $\tau\tau+jets$ channel, we generate the indirect production events with a cut at the generator level on the invariant mass of the $\tau\tau$ pair, $M(\tau,\tau)>140$ GeV to trim the overwhelmingly
large background events coming from the $Z$-boson peak. Similarly, in the $\tau+\slashed{E}_{\rm T}+jets$ channel, we apply a strong transverse mass cut, $m_{\rm T}(\tau,\nu)>250$ GeV at the generator level in order to suppress the large SM $W$ contribution. Now, because of the destructive nature of the interference term, there is a cancellation between the interference and the pure BSM contributions. However, even after avoiding the $Z$ or $W$ boson mass peaks, the SM contribution remains large and hence, $\sg_{\rm SM}> \sg^{excl}_{pp\to xy}$ [see Eq.~\eqref{eq:intf}]. In other words, once we include $\mc S_1$, the cross sections of the exclusive $pp\to\tau\tau,\tau\nu$ processes are lower than the 
expected SM prediction. 
\item
We define the efficiency for interference as,
\begin{eqnarray}
\varepsilon_{\times} = &&\frac{1}{\lt(-\lm^2\sg_{\times}\rt)}\Big\{\varepsilon_{xy}^{excl}\times\sg^{excl}_{pp\to xy}\nonumber\\
&&\quad\quad\hspace{1cm}-\ \varepsilon_{\rm SM}\times \sg_{\rm SM}-\ \varepsilon_{t}\times \lm^4 \sg_{t}\Big\},\quad\label{eq:epsilonX}
\end{eqnarray}
where $\varepsilon_{xy}^{excl}$, $\varepsilon_{\rm SM}$ and $\varepsilon_{t}$ are the efficiencies for the total exclusive $pp\to\tau\tau$ or $\tau \nu$ events, pure SM contribution and $t$-channel $\mc S_1$ exchange contribution, respectively. Notice, since both the numerator and the denominator in Eq.~\eqref{eq:epsilonX} are negative, $\varepsilon_{\times}$ is positive (Tables~\ref{tab:tautau} and
\ref{tab:taunu}). As we have already indicated earlier, in this scenario, the number of surviving events coming from the interference term is larger than that of all other LQ processes put together for both 
the $\tau\tau+jets$ and $\tau+\slashed{E}_{\rm T}+jets$ channels once we apply the selection cuts (mentioned in the last section). As a result, the number of predicted events in both the channels reduce when we include $\mc S_1$. 

\end{enumerate} 

\begin{figure*}[t]\vspace{-1.5ex}
\subfloat[Scenario-I]{\includegraphics[width=0.48\textwidth]{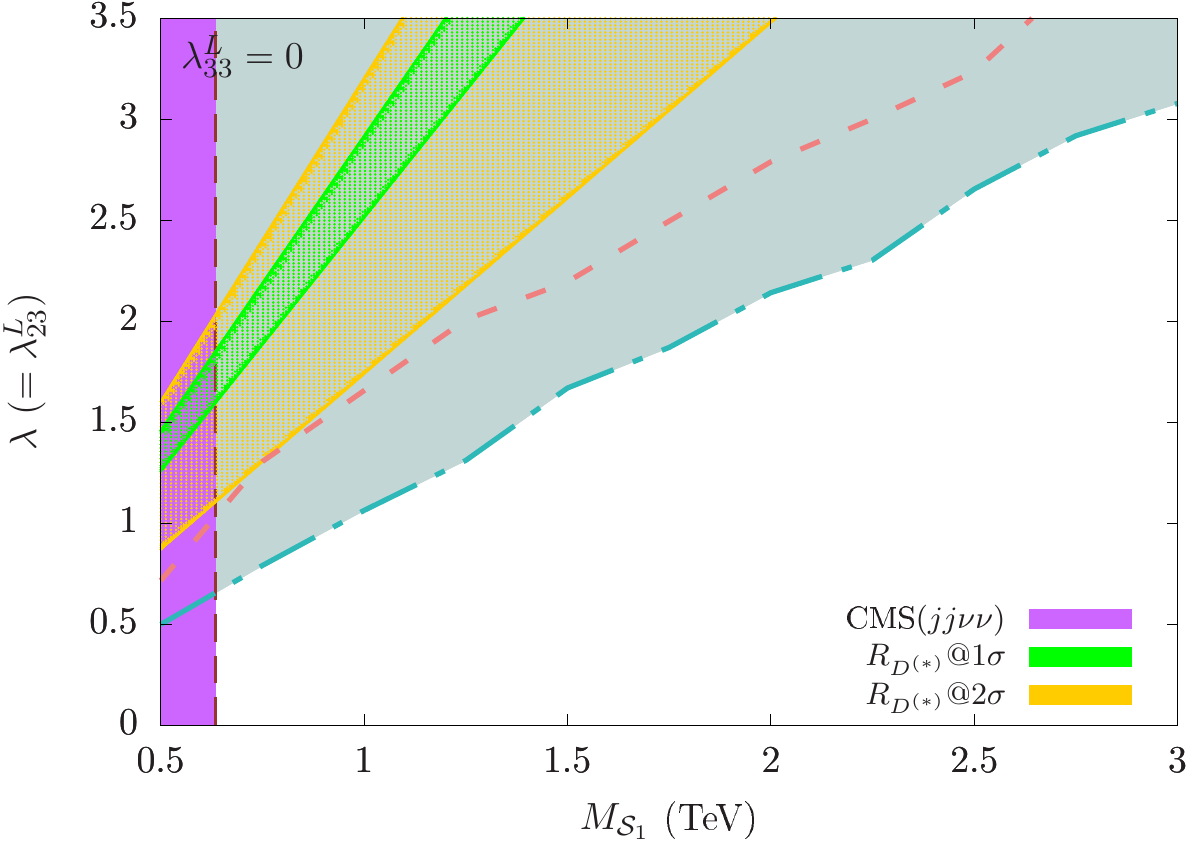}\label{fig:sce1}}\hfill
\subfloat[Scenario-II]{\includegraphics[width=0.48\textwidth]{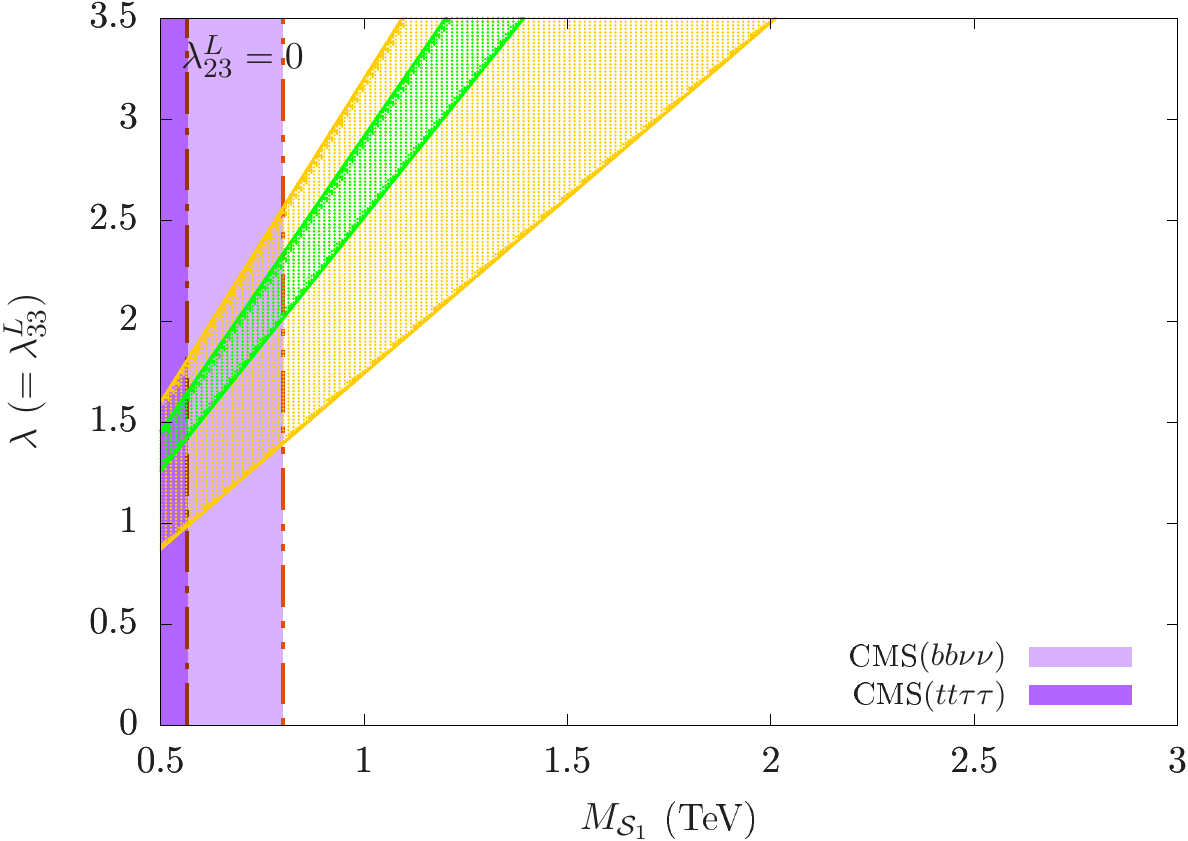}\label{fig:sce2}}\\
\subfloat[Scenario-III]{\includegraphics[width=0.48\textwidth]{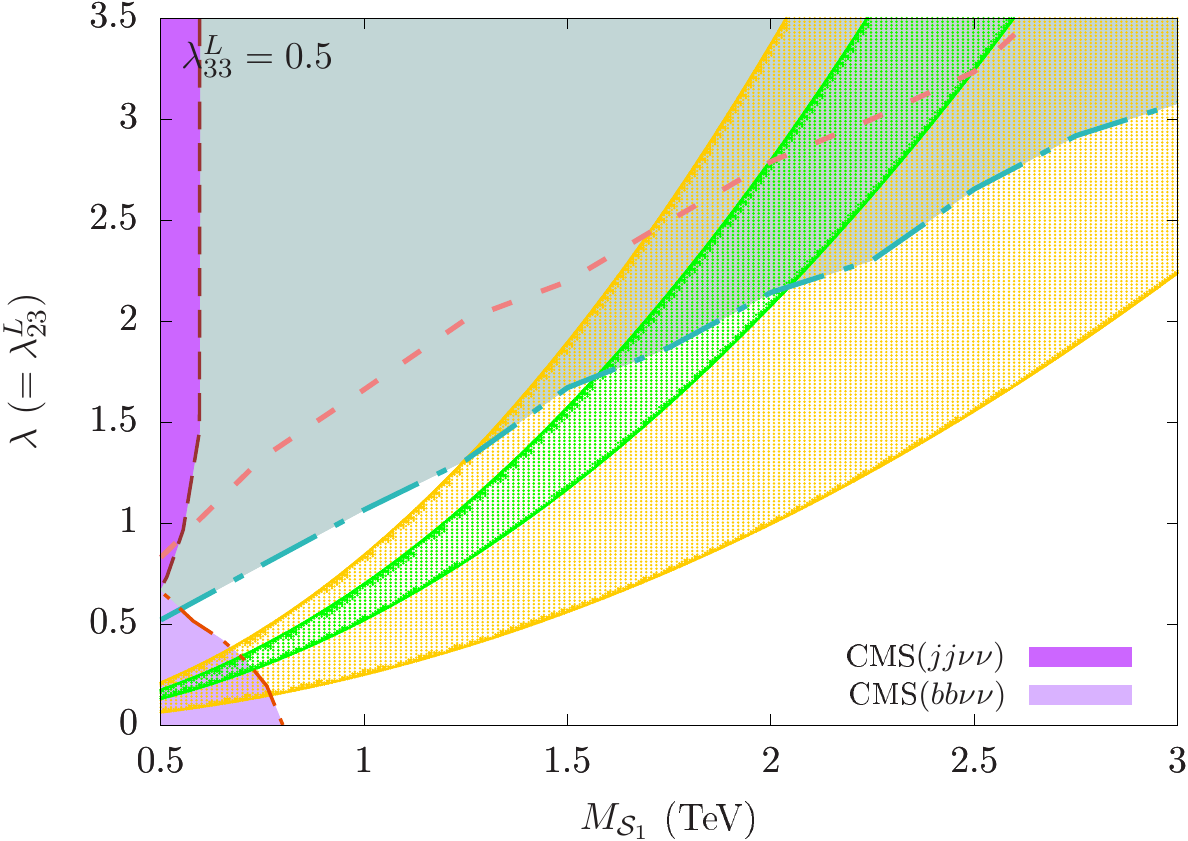}\label{fig:sce3a}}\hfill
\subfloat[Scenario-III]{\includegraphics[width=0.48\textwidth]{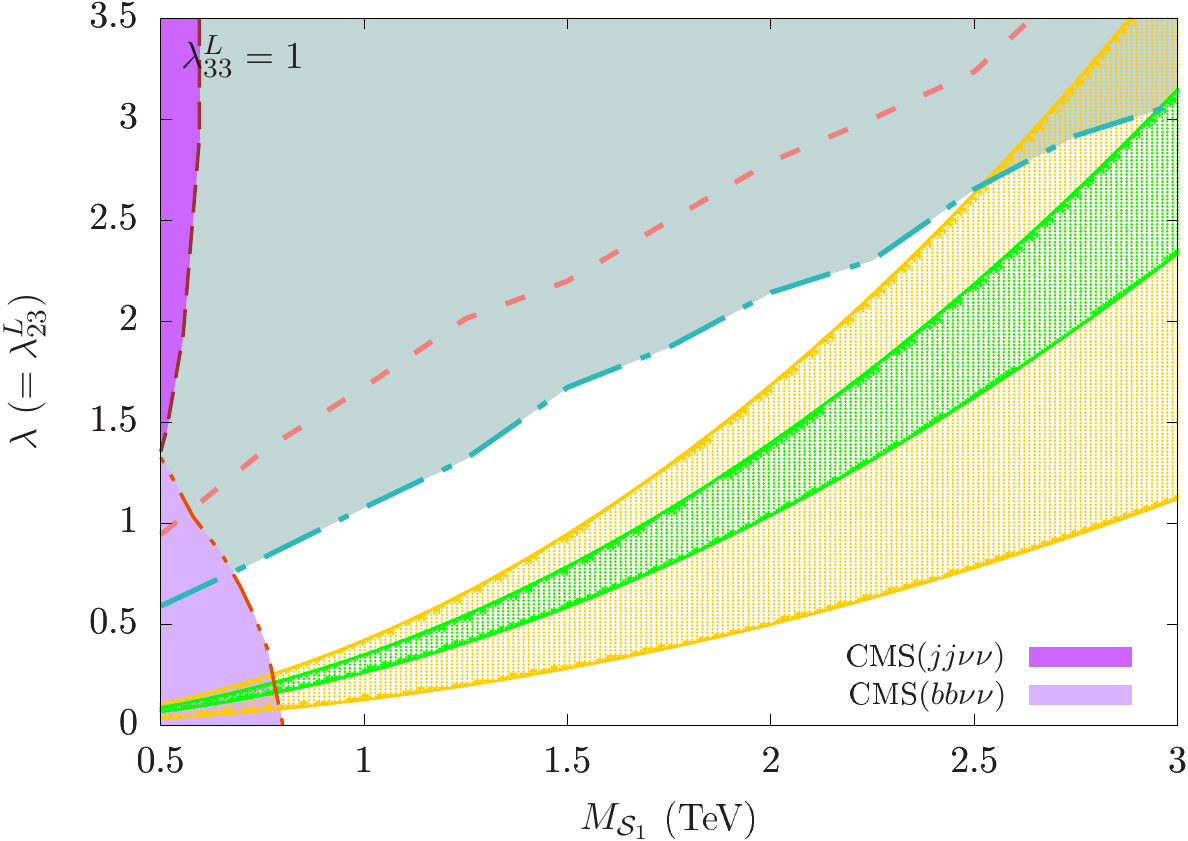}\label{fig:sce3b}}
\caption{The $95$\% CL ($2\sg$) exclusion limits from the LHC in the $M_{\mc S_1}-\lambda$ plane for the three scenarios in the minimal model with $\mc S_1$ [Eq.~\eqref{eq:Lcompact}] and the preferred regions by the $R_{D^{(*)}}$ anomalies with (a) $\lm=\lm^L_{23}$, $\lm^L_{33}=0$ (Scenario-I), (b) $\lm^L_{23}=0$, $\lm=\lm^L_{33}$ (Scenario-II),
(c) $\lm=\lm^L_{23}$, $\lm^L_{33}=0.5$ (Scenario-III) and (d) $\lm=\lm^L_{23}$, $\lm^L_{33}=1$ (Scenario-III).
The blue and red dashed lines show the $2\sg$ exclusion limits obtained by recasting the ATLAS
$\tau\tau$~\cite{Aaboud:2017sjh} and $\tau\nu$~\cite{Aaboud:2018vgh} resonance search data, respectively. The grey shaded region is the union of the regions excluded in these two channels. The excluded mass ranges from the direct (pair production) searches are shown with shades of purple -- in Scenario-I the limit [$635$ GeV, deep purple in (a)]  comes from the CMS search for $jj+\slashed E_{\rm T}$~\cite{Sirunyan:2018kzh},  in Scenario-II, from the CMS $tt\tau\tau$ ~\cite{Sirunyan:2018nkj} [$565$ GeV, shown with deep purple in (b)] and $bb+\slashed E_{\rm T}$ searches~\cite{Sirunyan:2018kzh} [$800$ GeV, shown with light purple in (b)]. We recast the pair production limit on Scenario-II from the CMS $bb+\slashed E_{\rm T}$ search~\cite{Sirunyan:2018kzh} data (i.e., the stronger one) in the Scenario-III plots [shown with light purple in (c) and (d)].  However, since with varying $\lm^L_{23}$, $Br\lt(\mc S_1\to b\nu\rt)$ varies, the limit changes with $\lm^L_{23}$ -- it decreases as $\lm^L_{23}$ increases. Similarly, the pair production limit from the CMS search for $jj+\slashed E_{\rm T}$~\cite{Sirunyan:2018kzh} on Scenario-I is recast for Scenario-III [shown with deep purple in (c) and (d)]. The green and yellow bands show the regions favoured by the $R_{D^{(*)}}$ anomalies within $1\sg$ and $2\sg$, respectively. We have used Eqs.~\eqref{eq:rds}--\eqref{eq:cv3} to obtain these. }
\label{fig:RDexclu}
\end{figure*}

This reduction in expected number of events causes difficulty in directly recasting the exclusion limits by rescaling the efficiencies as we did earlier in Refs.~\cite{Mandal:2015vfa,Mandal:2016csb}. Instead, we use the observed $m_{\rm T}^{tot}$ 
distribution from  the $\tau\tau$ search in Ref.~\cite{Aaboud:2017sjh} (from the $b$-veto category in the $\tau_{had}\tau_{had}$ and the $\tau_{lep}\tau_{had}$ modes. For consistency, we also apply the same $b$-veto on our events in these modes.) and the $m_{\rm T}$ distribution from the $\tau\nu$ search in Ref.~\cite{Aaboud:2018vgh} to perform a $\chi^2$ test. For that we bin the signal events passing through the basic signal selection criteria following the experimental distributions. For both the $\tau\tau$ and $\tau\nu$ channels, we define the test statistic as
\begin{eqnarray}
\chi^2&=& \displaystyle\mathlarger{\mathlarger{\sum}}_i \lt[\frac{ N_{\rm T}^i- N_{\rm D}^i }{\Dl N^i}\rt]^2,
\label{eq:chisq}
\end{eqnarray}
where the sum runs over all the bins. Here, $N_{\rm T}^i$ and $N_{\rm D}^i$ are the number of expected or the Monte Carlo (MC) simulated theory events and the number of observed events (data) in the $i^{\rm th}$ bin, respectively. The total simulated events in the $i^{\rm th}$ bin is obtained by,
\begin{eqnarray}
N_{\rm T}^i &=& N_{\mc S_1}^i + N_{\rm BG}^i\nn\\
&=& \lt[N_{p}+N^{incl}_s+N_t-N_{\times}\rt]^i + N_{\rm BG}^i,\quad
\end{eqnarray}
where  $N_{\mc S_1}^i$, $N_{\rm BG}^i$ are MC signal events and the SM background events in the $i^{\rm th}$ bin and 
$N_{p}$, $N^{incl}_s$, $N_t$ and $N_{\times}$ are the signal events from the pair production, the total inclusive single production, the pure BSM term of the $t$-channel $\mc S_1$ interchange and the interference contribution, respectively. For the error $\Dl N^i$ in the denominator of Eq.~\eqref{eq:chisq}, we use the total uncertainty,
\begin{align}
\Dl N^i = \sqrt{\lt(\Dl N^i_{\rm Stat}\rt)^2+\lt(\Dl N^i_{\rm Syst}\rt)^2}\ ,
\end{align}
where $\Dl N^i_{\rm Stat}=\sqrt{N^i_{\rm D}}$ and we assume 
$\Dl N^i_{\rm Syst}=\dl^i\times N^i_{\rm D}$. We extract $N_{\rm D}^i$ and $N_{\rm BG}^i$ from \textsc{HEPData}~\cite{Maguire:2017ypu}. To be conservative, we include a uniform 10\% systematic uncertainty (i.e., $\dl^i=0.1$) for all bins. Even if the actual systematic uncertainties are lower, it would not alter our results too much as the statistical uncertainties dominate in the error computations. To avoid spurious exclusions, we reject bins with $N^i_{\rm D}\leq 2$. 

We find that the SM provides a very good fit to the data in both the channels. We obtain the 
minimum value of $\chi^2=\chi^2_{min}$ and the corresponding value of $\lm=\lm_{min} \lt(\geq0\rt)$ for some benchmark values of $M_{\mc S_1}$ between $0.5$ TeV and $3$ TeV by varying $\lm$ in each case.\footnote{
It is interesting to note that for some benchmark masses, the $\chi^2/d.o.f.$ is slightly improved than the SM fit. Therefore, one could say that the presence of ${\mc S_1}$ is slightly favoured by the data. However, the improvement is marginal and hence not important statistically.} For every $M_{\mc S_1}$, we find the exclusion upper limit (UL) on $\lm$ by finding the boundaries of $1\sg$ and $2\sg$ confidence intervals in $\chi^2$. Since, for every benchmark $M_{\mc S_1}$, we vary only $\lm$ (i.e., effectively one variable), the $1\sg$ and $2\sg$ confidence level (CL) UL on $\lm$ will be given by $\lm$'s for which $\Dl\chi^2=1$ and $\Dl\chi^2=4$, respectively. Here, $\Dl\chi^2$ is defined as 
$\Dl\chi^2=\chi^2-\chi^2_{min}$. 

In Fig.~\ref{fig:lmexclu}, we show $1\sg$ and $2\sg$ CL UL on $\lm$ in Scenario-I from the $\tau\tau$ and $\tau\nu$  resonance data. We see that the $\tau\tau$ data gives stronger limit on $\lm$ for the entire range of $M_{\mc S_1}$. The limits from the CMS pair production search in the $2j+\slashed E_{\rm T}$ channel~\cite{Sirunyan:2018kzh} are shown in Fig.~\ref{fig:sce1}. We have obtained the pair production limit by simply rescaling the $\sg^{{\rm pp}\to {\rm LQ}_{\rm s}{\rm LQ}_{\rm s}}_{\rm theory, NLO}$ line from the first plot of Fig.~3 of Ref.~\cite{Sirunyan:2018kzh} by the square of $\displaystyle Br\left(\mc S_1\to s\nu\rt)$ and finding its new intersection with the observed limit. The intersection gives the lower limit on $M_{\mc S_1}$ ($635$ GeV) in Scenario-I that is independent of $\lm$. In the same plot we have also shown the regions favoured by the $R_{D^{(*)}}$ anomalies within $1\sg$ and $2\sg$, respectively. We have used Eq.~\eqref{eq:rds} to compute the corrections to the $R_{D^{(*)}}$ observables in Scenario-I and HFLAV averages \cite{Amhis:2016xyh} to obtain these regions. We see that the LHC data is not only sensitive to the parameters in Scenario-I, it has effectively ruled out Scenario-I as a possible explanation for the $R_{D^{(*)}}$ anomalies. Even a heavy $\mc S_1$ will not work.

Ideally, to do a proper recast of the pair production search result, one should consider the contribution of other $\lm$ dependent production processes (like the inclusive single production) to $2j+\slashed E_{\rm T}$ final states (as we have demonstrated the procedure in Ref.~\cite{Mandal:2015vfa}). Then one would get a mass dependent limit on the coupling from the pair production too. However, the limits obtained on $\lm$ are weaker than those shown here. Hence, in this paper we do simple recast of all the pair production searches for simplicity (even for Scenario-II).
\subsection{Bounds on Scenario-II}
\noindent
In this scenario, the single production cross sections are negligible. For example, the total cross section for $pp\to t\tau\nu$ via a one TeV $\mc S_1$ is just about $\sim 2$ fb (assuming $\lm^L_{33}=1$). One has to go to high luminosity to probe these signatures. Here, we show the pair production limits on Scenario-II in Fig.~\ref{fig:sce2}. The limits are obtained by simple rescaling of the ones obtained by the CMS $tt\tau\tau$ ~\cite{Sirunyan:2018nkj} and $bb+\slashed E_{\rm T}$ searches~\cite{Sirunyan:2018kzh} (both of these searches assume unit branching fraction in the respective searched channels) just like we did in Scenario-I. In Scenario-II, the limits obtained from the $tt\tau\tau$ and $bb+\slashed E_{\rm T}$ data are $565$ GeV and $800$ GeV, respectively. Here we see that to explain the $R_{D^{(*)}}$ anomalies in this scenario, one needs $M_{\mc S_1}> 800$ GeV with pretty high $\lm^L_{33}$ ($\gtrsim 1.5$).

\subsection{Bounds on Scenario-III}
\noindent
The limits on Scenario-III are shown for two benchmark values of $\lm^L_{33}$ ($0.5$ and $1.0$) in Figs.~\ref{fig:sce3a} \& \ref{fig:sce3b}. Like before, the grey shaded areas show the regions excluded by the ATLAS $\tau\tau$ and $\tau\nu$  resonance data. Unlike the indirect production, the pair and the inclusive single production contributions depend on the branching fractions in the $\tau j$ and $\nu j$ modes (remember that $j$ is an untagged jet, i.e., it could mean a light jet or a $b$-jet) which, in turn, depend on $\lm^L_{33}$. However, for large $M_{\mc S_1}$, direct production cross sections become negligible compared to the indirect ones. Hence, only for $M_{\mc S_1} \lesssim 1$ TeV, we see some minor differences between the grey areas in  Figs.~\ref{fig:sce3a} \& \ref{fig:sce3b} and that in Fig.~\ref{fig:sce1}. When $\lm^L_{23}\to0$, Scenario-III tends towards Scenario-II. Hence, the pair production limit on Scenario-II obtained from the CMS $bb+\slashed E_{\rm T}$ search~\cite{Sirunyan:2018nkj} is repeated in these figures. Again, due to the change in $Br\lt(\mc S_1\to b\nu\rt)$ with $\lm^L_{23}$, the limit varies. The limit decreases as $\lm^L_{23}$ increases. The pair production limit from the CMS search for $jj+\slashed E_{\rm T}$ channel~\cite{Sirunyan:2018kzh} on Scenario-I is also recast for Scenario-III after correcting for the appropriate $Br\lt(\mc S_1\to s\nu\rt)$. As expected, Scenario-III has more freedom to accommodate the $R_{D^{(*)}}$ anomalies. In this case, one does not need very large couplings, for example $\lm^L_{23},\lm^L_{33}\approx 0.5$ with $M_{\mc S_1}\sim 1$ TeV would be good to explain the anomalies (though such a choice of parameters would be ruled out by other flavour or electroweak bounds if one strictly considers this scenario). However, as $\lm^L_{33}$ becomes smaller (i.e., Scenario-III tends towards Scenario-I), the $R_{D^{(*)}}$-favoured space get in tension with the exclusion limits, especially for high $M_{\mc S_1}$. Interestingly, here we see that even the pair production limits still allow a lighter than a TeV $\mc S_1$.

\subsection{Bounds on $\lm^R_{23}$}

\begin{figure}[t]
\includegraphics[width=0.48\textwidth]{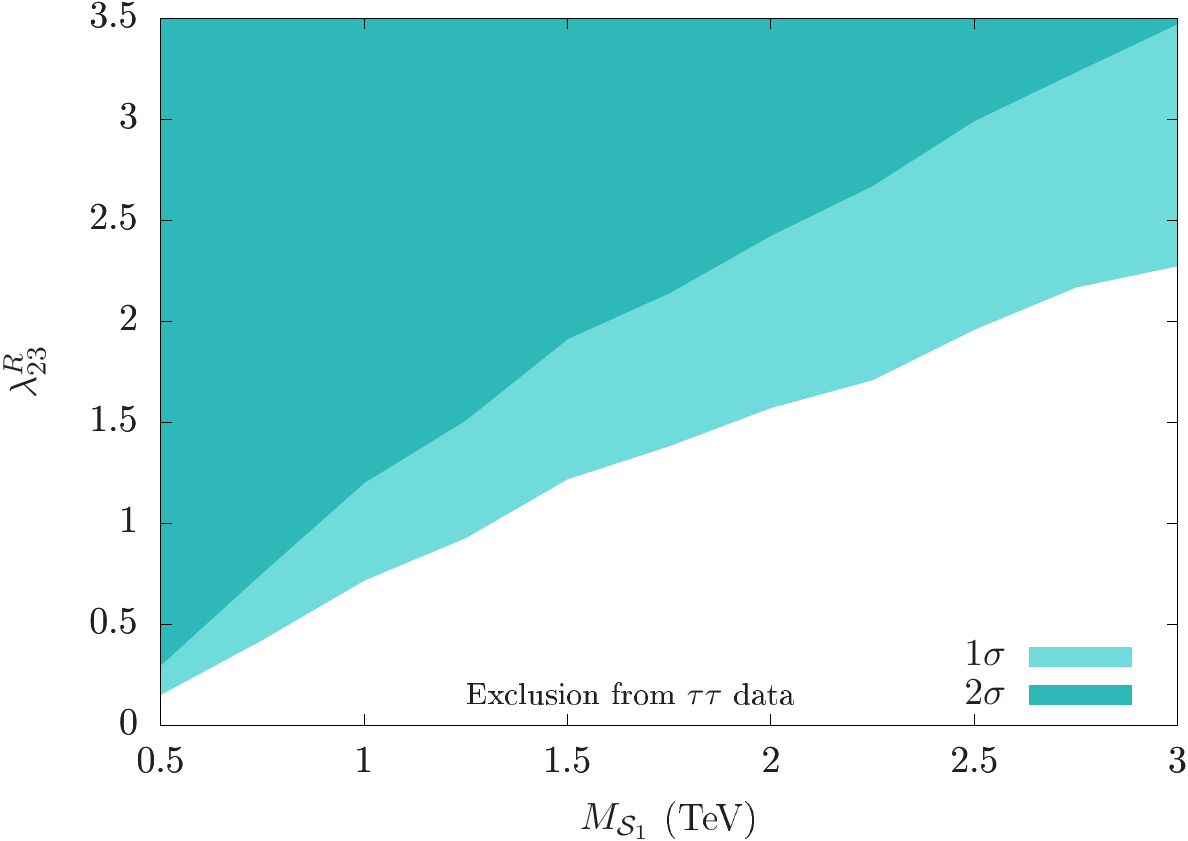}
\caption{The $1\sg$ and $2\sg$ CL exclusion limits on $\lm^R_{23}$ as functions of $M_{\mc{S}_1}$ using the ATLAS $\tau\tau$~\cite{Aaboud:2017sjh} resonance search data. The coloured regions are excluded. To obtain this plot we have assumed $\lm^L_{23},\lm^L_{3\al}=0$. We keep $\lm^R_{23}\leq3.5$ to ensure $\displaystyle\lt(\lm^R_{23}\rt)^2/4\pi<1$.}
\label{fig:LM23Rexclu}
\end{figure}

As indicated in Section~\ref{sec:mod}, for completeness we also display the limits on $\lm^R_{23}$ in the $M_{\mc S_1}-\lm^R_{23}$ plane in Fig.~\ref{fig:LM23Rexclu}. To obtain this plot, we set $\lm^L_{23},\lm^L_{3\alpha}=0$ in Eq.~\eqref{eq:Lcompact}, and, as a result, $\mc S_1$ can no longer couple with a neutrino (hence, this coupling alone cannot resolve the $R_{D^{(*)}}$ anomalies). We consider the ATLAS $\tau\tau$ resonance search data to obtain limits as it can produce only $\tau\tau+jets$ signature. Like in Scenario-I (Fig.~\ref{fig:lmexclutautau}), the (destructive) interference of the $t$-channel LQ exchange in $cc\to\tau\tau$ with the SM $cc\to Z\to\tau\tau$ plays the dominant role in setting the limits. The limits slightly differ from ones shown in Fig.~\ref{fig:lmexclutautau}. 
In the SM, the $Z$ boson couples weakly to a right handed $\tau$ than it does to a left handed one. Hence, the limits on $\lm^R_{23}$ are lower than $\lm^L_{23}$ for $M_{\mc S_1}\gtrsim 1$ TeV. Like before, the pair and the inclusive single productions play some roles in determining the exclusion limits for low $M_{\mc S_1}$. The pair and single production cross sections are, however, unaffected by the shift $\lm^L_{23}\to\lm^R_{23}$ except now $\mc S_1$ only decays to $c\tau$. Hence, the limits on $\lm^R_{23}$ is slightly stronger that those on $\lm^L_{23}$ for $M_{\mc S_1}\lesssim 1$ TeV.

\section{Summary and Conclusions}
\label{conclu}
\noindent
In this paper, we have studied the LHC signatures of a simple extension of the SM with a single charge $-1/3$ scalar LQ, denoted as $\mc S_1$, that can address the semileptonic $B$-decay anomalies observed in the $R_{D^{(*)}}$ observables. The possibility that such a LQ can address the $R_{D^{(*)}}$ anomalies has been discussed earlier in the literature. Here, however, our motivation is  
to investigate whether the LHC can give competitive bounds on the parameter spaces of such extensions. 

We have identified some minimal scenarios, where the model can be specified with just two new parameters -- the mass of LQ and a new coupling which, normally, is expected to be large to accommodate the $R_{D^{(*)}}$ anomalies. To explain the observed $R_{D^{(*)}}$ anomalies within the simple model, we need two nonzero couplings -- $b\nu \mc S_1$ and $c\tau \mc S_1$. In the minimal scenarios, one of these two couplings is generated from the other via quark-mixing.

In one minimal scenario, which we call as Scenario-I, $\mc{S}_1$ has large cross-generation coupling $\lm_{23}^L$ that connects second generation quarks and third generation leptons. In the other minimal scenario (Scenario-II), it couples largely with quarks and leptons from the third generation with strength $\lm_{33}^L$. For completeness, we consider a hybrid scenario (Scenario-III) where both of these above couplings are nonzero.

From a collider perspective, a novel and interesting aspect in Scenario-I and III is that they allow production of $\mc S_1$ at the LHC through the $s$- and $c$-quark initiated processes. This is unlike Scenario-II where $\mc S_1$ is basically a third generation LQ and is produced either in the gluon or the $b$-quark initiated processes (the couplings that are generated solely by quark-mixing are too small to play any noticeable role at the LHC). A large $\lm^L_{23}$ enhances the single production cross sections of $\mc S_1$ and also gives rise to a significant number of nonresonant $\tau\tau$ ($\tau\nu$) events through the $t$-channel $\mc S_1$ exchange $cc\to \tau\tau$ ($cs\to\tau\nu$) process. 
It would lead to other interesting signatures like $\tau\tau/\tau\nu~+$ (light) $jets$ which are yet to be searched for experimentally. 
Earlier, Ref.~\cite{Dumont:2016xpj} had considered $c$-quark initiated production. However, the scenario they considered had both $\lm^L_{3\alpha}$ and $\lm^R_{23}$ nonzero but $\lm^L_{23}=0$ [see Eq.~\eqref{eq:Lcompact}] and hence is different from Scenario-I, -II or -III.

Here, we have used the latest $Z^\prime$ and $W^\prime$ resonance search data from the ATLAS collaboration through the $Z'\to \tau\tau$~\cite{Aaboud:2017sjh} and 
$W^\prime\to\tau\nu$~\cite{Aaboud:2018vgh} channels to put bounds on Scenario-I and III. 
We have found that the indirect production processes strongly interfere with the similar SM processes in a destructive manner. The interference gives the dominant effect in the estimation of exclusion limits in Scenario-I and III for order one $\lm^L_{23}$. This destructive nature of the interference leads to a reduction of total number of expected 
SM events in the $pp\to\tau\tau$ or $pp\to\tau\nu$ processes. Because of this, we have performed a $\chi^2$ test using the experimentally obtained transverse mass distributions to derive exclusion limits on the $\lm^L_{23}-M_{\mc S_1}$ plane. (Previously, in Refs.~\cite{Faroughy:2016osc,Angelescu:2018tyl} where the $\tau\tau$ or $\tau\nu$ search data were used to obtain the exclusion limits on the LQ parameter space, the interference contribution was not considered.) In addition to the indirect production, we have included the inclusive single and pair production contributions systematically in the exclusion limit estimations from the $\tau\tau$ or $\tau\nu$ search data. We have found that the inclusive single production contributions, although small compared to the indirect production, leads to visible effects in the exclusion limits especially
for low $M_{\mc{S}_1}$.

The limits that we have obtained are realistic and proper since we systematically consider the indirect (including the interference contributions) and direct LQ productions in our analysis. We have found that the latest LHC $\tau\tau$ or $\tau\nu$ resonance search data is powerful enough to constrain the LQ parameter space in Scenario-I and III. In fact, it practically rules out the entire region favoured by the $R_{D^{(*)}}$ anomalies in Scenario-I. This is possible as, unlike the direct pair production search data, it gives a $\lm^L_{23}$ dependent exclusion boundary that goes up to large values of $M_{\mc S_1}$. For small $\lm^L_{33}$ in Scenario-III (when it comes closer to Scenario-I), the exclusion limits are in tension with the $R_{D^{(*)}}$-favoured parameters, but with large $\lm^L_{33}$ (when Scenario-III moves towards Scenario-II), the tension goes away. In Scenario-II, the strongest limit comes from the direct pair production search by CMS in the $2b+\slashed E_{\rm T}$ channel. This excludes $M_{\mc S_1} < 800$ GeV in this scenario. Similarly, the pair production search by CMS in the $2j+\slashed E_{\rm T}$ channel excludes $M_{\mc S_1} < 635$ GeV in Scenario-I.

As we have clearly mentioned before, the three scenarios we have considered are simplistic and, 
on their own, would have a hard time facing other flavour or precision electroweak bounds~\cite{Hiller:2016kry,Cai:2017wry,Angelescu:2018tyl} if one looks beyond the $R_{D^{(*)}}$ anomalies. In fact, not only with $\mc S_1$, all single 
LQ solutions to the flavour anomalies get in conflict with some bound or other (see e.g. Refs.~\cite{Angelescu:2018tyl,Bansal:2018nwp}). One has to make additional theoretical constructions to avoid the tension. However, even then, the limits we have obtained would still be meaningful as long as the couplings from Eq.~\eqref{eq:Lcompact} 
are not negligible. It is easy to see that the pair production bounds would be applicable in any extension of the $\mc S_1$ model. 
However, since in any channel the pair production contribution is sensitive to the corresponding 
branching fraction, the limits from pair production channels have to be rescaled with square of the respective 
branching fractions. 

The bounds we show in Fig.~\ref{fig:lmexclu}, come predominantly from the interference of $t$-channel LQ exchanges 
with the SM background. The interference depends only on the coupling involved, but (practically) not on 
the total width of $\mc{S}_1$ (hence, the branching fractions). As a result, these limits would be applicable in 
any extension of Scenario-I as long as there is no additional significant interference in these processes. For small 
$M_{\mc S_1}$ ($\lesssim 1$ TeV), the limits do get some noticeable contributions from inclusive single 
productions (which depends on the branching fractions) and will vary for different total widths but this 
difference would not be drastic.  A similar argument would also hold for the bounds shown in Fig.~\ref{fig:LM23Rexclu} 
in extensions with nonzero $\lm^R_{23}$. For example, in the scenario considered in Ref.~\cite{Dumont:2016xpj}, where 
$\lm^L_{3\alpha}$ and $\lm^R_{23}$ are nonzero but $\lm^L_{23}=0$, the pair production limits from 
Fig.~\ref{fig:sce2} (after rescaling for the branching fractions) and the limits from Fig.~\ref{fig:LM23Rexclu} 
would be applicable (for light $\mc S_1$, the limits on $\lm^R_{23}$ would be slightly off unless one corrects 
them for the appropriate branching fractions). Similarly, even if one considers all the three couplings to be nonzero, 
it is possible to obtain approximate bounds easily on a combination of $\lm^L_{23}$ and $\lm^R_{23}$ (namely, $\sqrt{(\lm^L_{23})^2+(\lm^R_{23})^2}$) by adopting the 
limits from the $\tau\tau$ data~\cite{Aydemir:2019ynb}. 

We can also consider the example mentioned earlier in the introduction from Ref.~\cite{Crivellin:2017zlb} 
where, in addition to the $\mc S_1$, a weak-triplet LQ, $\mc S_3$
 is introduced to cancel the contribution of $\mc S_1$ to $b\to s\nu\nu$ while $R_{D^{(*)}}$ gets contribution from both. 
For this, one needs the mass of the charge $1/3$ component of $\mc S_3$ (let us call it $\mc S^{1/3}_3$) to be
the same as $M_{\mc S_1}$.
If the $\mc S_3$ mass matrix is such that the other components of  
$\mc S_3$ (namely $\mc S^{4/3}_3$ 
and $\mc S^{2/3}_3$) are much heavier than $\mc S^{1/3}_3$, we can obtain rough limits on this scenario easily. Since the SM $cc\to\tau\tau$ 
process would now interfere with both $\mc S_1$ and $\mc S^{1/3}_3$ mediated processes similarly, the limits 
from $\tau\tau$ 
data have to be interpreted in terms of $\lm/\sqrt2$ where $\lm$ denotes the magnitude of the coupling strengths of both $\mc S_1$ and $\mc S^{1/3}_3$. 
Hence we see that it is possible to estimate limits from the LHC in various scenarios or models that contain the Lagrangian of Eq.~\eqref{eq:Lcompact} by adopting our results. If, however, in the $\mc S_3$-example, the other components also have masses comparable to $M_{\mc S_1}$ (so that they too would contribute to the $pp\to\ta\ta/\ta\nu$ processes significantly), one has to compute their contributions and follow our method explicitly to obtain the precise limits. The same can be said for models with significant additional contribution to the $pp\to\ta\ta/\ta\nu$ processes.

Finally, we note that the $\tau\tau$ or $\tau\nu$ resonance searches are not optimized for the nonresonant $t$-channel indirect production. As a result, in our recast, a large fraction of the signal events were lost. This can be seen in the small cut-efficiencies we have obtained. It is, therefore, important to make a dedicated search for LQs by optimizing cuts for the nonresonant indirect production including interference (either constructive or destructive) contribution in the signal definition.

\acknowledgments 
\noindent T.M. is supported by the INSPIRE Faculty Fellowship of the Department of Science and Technology (DST) under grant number IFA16-PH182 at University of Delhi. T.M. is also grateful to the Royal
Society of Arts and Sciences of Uppsala for financial support as a guest researcher at Uppsala University. S.M. acknowledges support from the Science and
Engineering Research Board (SERB), DST, India under grant number ECR/2017/000517. We thank R. Arvind Bhaskar for reading and commenting on the manuscript.

\bibliography{reference}{}
\bibliographystyle{JHEPCust}

\end{document}